\documentclass[a4paper, 11pt]{article} 
\usepackage{jheppub}\makeatletter\gdef\@fpheader{}\makeatother
\usepackage{mathtools, tabularx, subfigure, graphicx, cleveref}
\usepackage[expansion=false]{microtype}
\usepackage[dvipsnames]{xcolor}
\usepackage{youngtab}\Yboxdim 6.5pt\Ylinethick 0.4pt
\graphicspath{{Figures/}}

\def\a{A}
\def\b{B}
\def\c{C}
\def\d{D}
\def\e{E}
\def\f{F}
\def\g{G}
\def\p{P}
\def\q{Q}

\author[a]{Mudassar Sabir,}
\author[b,c]{Tianjun Li,}
\author[d]{Adeel Mansha,}
\author[e]{Xiao-Chuan Wang}

\affiliation[a]{Institute of Geophysics and Geomatics, China University of Geosciences, Wuhan, Hubei 430074, P. R. China}
\affiliation[b]{CAS Key Laboratory of Theoretical Physics, Institute of Theoretical Physics, Chinese Academy of Sciences, Beijing 100190, P. R. China}
\affiliation[c]{School of Physical Sciences, University of Chinese Academy of Sciences, Beijing, P. R. China}
\affiliation[d]{Department of Physics, Zhejiang Normal University, Jinhua 321004, P. R. China}
\affiliation[e]{Department of Physics, Henan Normal University, Xinxiang, Henan, 453007, P. R. China}

\emailAdd{sabir@cug.edu.cn}
\emailAdd{tli@itp.ac.cn}
\emailAdd{adeelmansha@alumni.itp.ac.cn}
\emailAdd{xcwang@htu.edu.cn}

\keywords{}
\arxivnumber{2202.07048}

\begin{document}

\title{The Supersymmetry Breaking Soft Terms, and
 Fermion Masses and Mixings in the Supersymmetric Pati-Salam Model from Intersecting D6-branes}

\abstract{A supersymmetric Pati-Salam model with wrapping number equal to 5 has been constructed in Type IIA orientifolds on $\mathbf{T}^6/(\mathbb{Z}_2\times \mathbb{Z}_2)$ with intersecting D6-branes recently. In particular, the string-scale gauge coupling unification can be achieved due to the intermediate-scale vector-like particles from $\mathcal{N}=2$ sector. We calculate the supersymmetry breaking soft terms, and study the Standard Model (SM) fermion masses and mixings. There are nine pairs of Higgs doublets from $\mathcal{N}=2$ sector. Interestingly, we can explain the SM quark masses and mixings, as well as the charged leptons masses from three point and four-point Yukawa interactions. Moreover, we calculate the supersymmetry breaking soft terms in a previous model with gauge coupling unification since we find a typo in the previous study.}

\maketitle
\flushbottom
\section{Introduction}\label{sec:Intro}
One of the early goals of string phenomenology has been to successful embed the known particle physics
Standard Model (SM) in a certain Calabi-Yau compactification in string theory. The earliest attempts
were mainly focused on models from weakly coupled heterotic string theory with gauge group
$E_8 \times E_8$. Later, with the discovery of D-branes the gauge groups came to be understood
as the stacks of coincident D-branes with open strings connecting them in various fashions
in the context of type II string theory. However, getting down from ten dimensions to
positively curved (de Sitter) universe with observed three families of chiral fermions
where all moduli are stabilized is still a distant goal. Importantly, the scale of supersymmetry breaking
is still  unknown. The situation provides ample room for theorists to come up
with semi-realistic models of standard model starting from any of the ten dimensional heterotic string theory,
type II theories or F-theory in twelve dimensions.

In the SM the light fermions appear in chiral representations of the ${\rm SU}(3)_C\times {\rm SU}(2)_W \times {\rm U}(1)_Y$
gauge group such that all gauge anomalies are canceled. The simplest case of parallel D-branes
in flat space does not yield chiral fermions. One way to realize the chiral fermions is to place D-branes on orbifold singularities. Another way is to consider intersecting D-branes on generalized orbifolds called orientifolds. In addition to the discrete internal symmetries of the world-sheet theory, that are gauged in orbifold constructions, the products of internal symmetries with world-sheet parity reversal become also gauged in orientifolds.

In this paper we restrict ourselves to the study of intersecting D6-branes models from the perspective of IIA string theory. Prior to string theory, within the context of non-supersymmetric four dimensional GUT model building, there was no candidate gauge group where the three chiral families of standard model could be put in one irreducible representation without introducing the ``antifamilies'' of opposite chirality \cite{Witten:2002ei}. Models in type IIA string theory can achieve the family replication by the multiple intersections of intersecting D6-branes. D6-branes fill the 4-dimensional spacetime and have 3 extra dimensions along the compactified directions in IIA string theory. As the latter three extra dimensions are exactly equal to half of the number of the compactified dimensions, thus two generic D6-branes intersect at one point of the extra dimensions. This intersection is where the fields arising from open strings stretched between two different D6-branes live.

The volume of the cycles that the D-branes wrap around determines the four dimensional gauge couplings and the total internal volume yields the gravitational coupling. The cubic couplings such as the Yukawa couplings may be calculated from open world sheet instantons \textit{i.e.}, triangular fundamental worldsheets stretched between the three intersections where the three fields involved in the cubic coupling reside. This is of great advantage since open world sheet instanton effects are naturally suppressed with $\exp(-A_{ijk}\,T)$ where $A_{ijk}$ is the worldsheet area of the triangle bounded by the intersections $\{i,j,k\}$  and $T$ is the string tension. This exponential function takes care of the mass hierarchies and mixings of the fermions. The general flavor structure and selection rules for intersecting D-brane models has been investigated in \cite{Chamoun:2003pf, Higaki:2005ie}.

In the typical toroidal orientifold compactifications, not all of the fermions sit on the localized intersections on the same torus which results in the rank-1 problem of the Yukawa mass matrices. Later, a number of models were eventually found where the Yukawa mass matrices do not have rank-1 problem \cite{Cvetic:2004ui}. Recently in ref.~\cite{Li:2019nvi} some new supersymmetric Pati-Salam models from intersecting D6-branes on a $\mathbf{T}^6/(\mathbb{Z}_2\times \mathbb{Z}_2)$ orientifold in IIA string theory have been constructed using the methods of machine learning. Here, we discuss the phenomenology of a particular class of these newly found models where one of the wrapping numbers is 5.

The model exhibits approximate gauge coupling unification and contains nine Higgs fields from the $\mathcal{N}=2$ subsector. Despite more freedom, all standard model fermion masses cannot be exactly fitted in the simplest case where the Wilson fluxes are set to zero. We find two interesting solutions in the parametric space where either all quarks and the heaviest charged lepton or all leptons and the heaviest quark masses; can be fitted with the extrapolated values obtained from running RGEs up to the unification scale. Of course exact matching may be achieved by turning on fluxes or invoking higher dimensional operators. Adding possible contributions from the classical four-point interactions, we explicitly show exact matching of all SM fermion masses and mixings. We also discuss the F-term breaking of the supersymmetry and calculate the soft terms from supersymmetry breaking for the $u$-moduli dominant cases with and without the dilaton $s$. The $t$-moduli dominant case is left-out as there the soft terms are not independent of the Yukawa couplings.

This paper is organized as follows. In section~\ref{sec:orientifold}, we will extract minimal supersymmetric standard model from intersecting D6-branes on a $\mathbf{T^6/(\mathbb{Z}_2\times \mathbb{Z}_2)}$ orientifold. In section~\ref{sec:EFT} we discuss the 4-dimensional effective field theory and the relevant soft terms from supersymmetry breaking. In section~\ref{sec:Yukawa} we derive Yukawa couplings in intersecting D6-brane model on Type IIA $\mathbf{T^6/(\mathbb{Z}_2\times \mathbb{Z}_2)}$ orientifold. Utilizing the obtained Yukawa mass matrices we obtain the fermion masses and mixings for specific choices of the open and closed string-moduli (VEVs) in section~\ref{sec:Masses}. We then discuss possible corrections to fermion masses from higher-dimensional 4-point interactions in section~\ref{sec:4point}. Finally, we conclude in section~\ref{sec:conclusion}. The soft terms from a previously studied model having exact guage coupling unification are also computed in the appendix \ref{Appendix}.

\section{The Pati-Salam model building from $\mathbf{T^6/(\mathbb{Z}_2\times \mathbb{Z}_2)}$ orientifold} \label{sec:orientifold}

In the orientifold $\mathbf{T^6/(\mathbb{Z}_2\times \mathbb{Z}_2)}$, $\mathbf{T^6}$ is a product of three 2-tori with the orbifold group $(\mathbb{Z}_2\times \mathbb{Z}_2)$ has the generators $\theta$ and $\omega$ which are respectively associated with the twist vectors $(1/2,-1/2,0)$ and $(0,1/2,-1/2)$ such that their action on complex coordinates $z_i$ is given by,
\begin{eqnarray}
& \theta: & (z_1,z_2,z_3) \to (-z_1,-z_2,z_3), \nonumber \\
& \omega: & (z_1,z_2,z_3) \to (z_1,-z_2,-z_3). \label{orbifold}
\end{eqnarray}
Orientifold projection is the gauged $\Omega R$ symmetry, where $\Omega$ is world-sheet parity that interchanges the left- and right-moving sectors of a closed string and swaps the two ends of an open string as,
\begin{align}
\textrm{Closed}:  &\quad \Omega : (\sigma_1, \sigma_2) \mapsto (2\pi -\sigma_1, \sigma_2), \nonumber \\
\textrm{Open}:  &\quad  \Omega : (\tau, \sigma) \mapsto (\tau, \pi - \sigma) ,
\end{align}
and $R$ acts as complex conjugation on coordinates $z_i$. This results in four different kinds of orientifold 6-planes (O6-planes) corresponding to $\Omega R$, $\Omega R\theta$, $\Omega R\omega$, and $\Omega R\theta\omega$ respectively. These orientifold projections are only consistent with either the rectangular or the tilted complex structures of the factorized 2-tori. Denoting the wrapping numbers for the rectangular and tilted tori as $n_a^i[a_i]+m_a^i[b_i]$ and $n_a^i[a'_i]+m_a^i[b_i]$ respectively, where $[a_i']=[a_i]+\frac{1}{2}[b_i]$. Then a generic 1-cycle $(n_a^i,l_a^i)$ satisfies $l_{a}^{i}\equiv m_{a}^{i}$ for the rectangular 2-torus and $l_{a}^{i}\equiv 2\tilde{m}_{a}^{i}=2m_{a}^{i}+n_{a}^{i}$ for the tilted 2-torus such that $l_a^i-n_a^i$ is even for the tilted tori.

The homology cycles for a stack $a$ of $N_a$ D6-branes along the cycle $(n_a^i,l_a^i)$ and their $\Omega R$ images ${a'}$ stack of $N_a$ D6-branes with cycles $(n_a^i,-l_a^i)$ are respectively given as,
\begin{align}
[\Pi_a ]&=\prod_{i=1}^{3}\left(n_{a}^{i}[a_i]+2^{-\beta_i}l_{a}^{i}[b_i]\right), \nonumber \\
[\Pi_{a'}] &=\prod_{i=1}^{3}\left(n_{a}^{i}[a_i]-2^{-\beta_i}l_{a}^{i}[b_i]\right),
\end{align}
where $\beta_i=0$ or $\beta_i=1$ for the rectangular or tilted $i^{\rm th}$ 2-torus, respectively.
The homology three-cycles, which are wrapped by the four O6-planes, are given by
\begin{alignat}{2}
\Omega R : &\quad& [\Pi_{\Omega R}] &= 2^3 [a_1]\times[a_2]\times[a_3],  \nonumber\\
\Omega R\omega : && [\Pi_{\Omega R\omega}] &=-2^{3-\beta_2-\beta_3}[a_1]\times[b_2]\times[b_3],  \nonumber\\
\Omega R\theta\omega : && [\Pi_{\Omega R\theta\omega}] &=-2^{3-\beta_1-\beta_3}[b_1]\times[a_2]\times[b_3], \nonumber\\
\Omega R\theta : && [\Pi_{\Omega R \theta}] &=-2^{3-\beta_1-\beta_2}[b_1]\times[b_2]\times[a_3]. \label{orienticycles}
\end{alignat}
The intersection numbers can be calculated in terms of wrapping numbers as,
\begin{align}
I_{ab}&=[\Pi_a][\Pi_b] =2^{-k}\prod_{i=1}^3(n_a^il_b^i-n_b^il_a^i),\nonumber\\
I_{ab'}&=[\Pi_a]\left[\Pi_{b'}\right] =-2^{-k}\prod_{i=1}^3(n_{a}^il_b^i+n_b^il_a^i),\nonumber\\
I_{aa'}&=[\Pi_a]\left[\Pi_{a'}\right] =-2^{3-k}\prod_{i=1}^3(n_a^il_a^i),\nonumber\\
I_{aO6}&=[\Pi_a][\Pi_{O6}] =2^{3-k}(-l_a^1l_a^2l_a^3+l_a^1n_a^2n_a^3+n_a^1l_a^2n_a^3+n_a^1n_a^2l_a^3),\label{intersections}
\end{align}
where $k=\sum_{i=1}^3\beta_i$ and $[\Pi_{O6}]=[\Pi_{\Omega R}]+[\Pi_{\Omega R\omega}]+[\Pi_{\Omega R\theta\omega}]+[\Pi_{\Omega R\theta}]$.

\subsection{Constraints from tadpole cancellation and supersymmetry}\label{subsec:constraints}

Since D6-branes and O6-orientifold planes are the sources of Ramond-Ramond charges they are constrained by the Gauss's law in compact space implying the sum of D-brane and cross-cap RR-charges must vanishes \cite{Gimon:1996rq}
\begin{eqnarray}\label{RRtadpole}
\sum_a N_a [\Pi_a]+\sum_a N_a \left[\Pi_{a'}\right]-4[\Pi_{O6}]=0,
\end{eqnarray}
where the last terms arise from the O6-planes, which have $-4$ RR charges in D6-brane charge units. RR tadpole constraint is sufficient to cancel the ${\rm SU}(N_a)^3$ cubic non-Abelian anomaly while ${\rm U}(1)$ mixed gauge and gravitational anomaly or $[{\rm SU}(N_a)]^2 {\rm U}(1)$ gauge anomaly can be cancelled by the Green-Schwarz mechanism, mediated by untwisted RR fields \cite{Green:1984sg}.

Let us define the following products of wrapping numbers,
\begin{alignat}{4}
A_a &\equiv -n_a^1n_a^2n_a^3,   &\quad B_a &\equiv n_a^1l_a^2l_a^3,       &\quad     C_a &\equiv l_a^1n_a^2l_a^3,  &\quad   D_a &\equiv l_a^1l_a^2n_a^3, \nonumber\\
\tilde{A}_a &\equiv -l_a^1l_a^2l_a^3, & \tilde{B}_a &\equiv l_a^1n_a^2n_a^3, & \tilde{C}_a &\equiv n_a^1l_a^2n_a^3, & \tilde{D}_a &\equiv n_a^1n_a^2l_a^3.\,\label{variables}
\end{alignat}
Cancellation of RR tadpoles requires introducing a number of orientifold planes also called ``filler branes'' that trivially satisfy the four-dimensional ${\cal N}=1$ supersymmetry conditions. The no-tadpole condition is given as,
\begin{align}
 -2^k N^{(1)}+\sum_a N_a A_a&=-2^k N^{(2)}+\sum_a N_a B_a= \nonumber\\
 -2^k N^{(3)}+\sum_a N_a C_a&=-2^k N^{(4)}+\sum_a N_a D_a=-16,\,
\end{align}
where $2 N^{(i)}$ is the number of filler branes wrapping along the $i^{\rm th}$ O6-plane. The filler branes belong to the hidden sector $USp$ group and carry the same wrapping numbers as one of the O6-planes as shown in table~\ref{orientifold}. $USp$ group is hence referred with respect to the non-zero $A$, $B$, $C$ or $D$-type.

\begin{table}[t]
\caption{The wrapping numbers for four O6-planes.}
\begin{center}
\begin{tabular}{|c|c|c|}
\hline
  Orientifold Action & O6-Plane & $(n^1,l^1)\times (n^2,l^2)\times (n^3,l^3)$\\
\hline
    $\Omega R$& 1 & $(2^{\beta_1},0)\times (2^{\beta_2},0)\times (2^{\beta_3},0)$ \\
\hline
    $\Omega R\omega$& 2& $(2^{\beta_1},0)\times (0,-2^{\beta_2})\times (0,2^{\beta_3})$ \\
\hline
    $\Omega R\theta\omega$& 3 & $(0,-2^{\beta_1})\times (2^{\beta_2},0)\times (0,2^{\beta_3})$ \\
\hline
    $\Omega R\theta$& 4 & $(0,-2^{\beta_1})\times (0,2^{\beta_2})\times (2^{\beta_3},0)$ \\
\hline
\end{tabular}
\end{center}
\label{orientifold}
\end{table}
Preserving ${\cal N}=1$ supersymmetry in four dimensions after compactification from ten-dimensions restricts the rotation angle of any D6-brane with respect to the orientifold plane to be an element of ${\rm SU}(3)$, i.e.
\begin{equation}
\theta^a_1 + \theta^a_2 + \theta^a_3 = 0 \mod 2\pi ,
\end{equation}
with $\theta^a_j = \arctan (2^{- \beta_j} \chi_j l^a_j/n^a_j)$. $\theta_i$ is the angle between the $D6$-brane and orientifold-plane in the $i^{\rm th}$ 2-torus and $\chi_i=R^2_i/R^1_i$ are the complex structure moduli for the $i^{\rm th}$ 2-torus.
${\cal N}=1$ supersymmetry conditions are given as,
\begin{eqnarray}
x_A\tilde{A}_a+x_B\tilde{B}_a+x_C\tilde{C}_a+x_D\tilde{D}_a=0,\nonumber\\
\frac{A_a}{x_A}+\frac{B_a}{x_B}+\frac{C_a}{x_C}+\frac{D_a}{x_D} < 0, \label{susyconditions}
\end{eqnarray}
where $x_A=\lambda,\; x_B=2^{\beta_2+\beta_3}\cdot\lambda /\chi_2\chi_3,\; x_C=2^{\beta_1+\beta_3}\cdot\lambda /\chi_1\chi_3,\; x_D=2^{\beta_1+\beta_2}\cdot\lambda /\chi_1\chi_2$.

Orientifolds also have discrete D-brane RR charges classified by the $\mathbb{Z}_2$ K-theory groups, which are subtle and invisible by the ordinary homology~\cite{Witten:1998cd, Cascales:2003zp, Marchesano:2004yq, Marchesano:2004xz}, which should also be taken into account~\cite{Uranga:2000xp}. The K-theory conditions are,
\begin{eqnarray}
\sum_a \tilde{A}_a  = \sum_a  N_a  \tilde{B}_a = \sum_a  N_a  \tilde{C}_a = \sum_a  N_a \tilde{D}_a = 0 \textrm{ mod }4 \label{K-charges}~.~\,
\end{eqnarray}
In our case, we avoid the nonvanishing torsion charges by taking an even number of D-branes, {\it i.e.}, $N_a \in 2 \mathbb{Z}$.

\subsection{Particle spectrum}

\begin{table}[th]
\caption{General spectrum for intersecting D6-branes at generic angles, where ${\cal M}$ is the multiplicity, and $a_S$ and $a_A$ denote respectively the symmetric and antisymmetric representations of ${\rm U}(N_a/2)$. Positive intersection numbers in our convention refer to the left-handed chiral supermultiplets. }
\renewcommand{\arraystretch}{1.3}
\centering
\begin{tabular}{|c|c|}
\hline {\bf Sector} & \phantom{more space inside this box}{\bf
Representation}
\phantom{more space inside this box} \\
\hline\hline
$aa$   & ${\rm U}(N_a/2)$ vector multiplet  \\
       & 3 adjoint chiral multiplets  \\
\hline $ab+ba$   & $ {\cal M}(\frac{N_a}{2}, \frac{\overline{N_b}}{2})= I_{ab}(\yng(1)_{a},\overline{\yng(1)}_{b})$ \\
\hline $ab'+b'a$ & $ {\cal M}(\frac{N_a}{2}, \frac{N_b}{2})=I_{ab'}(\yng(1)_{a},\yng(1)_{b})$ \\
\hline $aa'+a'a$ &  ${\cal M} (a_S)= \frac 12 (I_{aa'} - \frac 12 I_{aO6})$ $\yng(2)$\\
                 &  ${\cal M} (a_A)= \frac 12 (I_{aa'} + \frac 12 I_{aO6}) $ $\yng(1,1)$\\
\hline
\end{tabular}
\label{tab:spectrum}
\end{table}

To have three families of the SM fermions, we need one torus to be tilted, which is chosen to be
the third torus. So we have  $\beta_1=\beta_2=0$ and $\beta_3=1$.
Several supersymmetric Pati-Salam models from intersecting D6-branes on a $\mathbf{T}^6/(\mathbb{Z}_2\times \mathbb{Z}_2)$ orientifold in IIA string theory were constructed in ref.~\cite{Cvetic:2004ui} up to the wrapping number 3. The phenomenology of such models up to the wrapping number of 3 was first studied in ref.~\cite{Chen:2007zu}. The general particle representations for intersecting D6-branes models at angles are shown in table~\ref{tab:spectrum}.

\begin{table}[h]
	\footnotesize
	\renewcommand{\arraystretch}{1.0}
	\caption{D6-brane configurations and intersection numbers in Model 16, and its MSSM gauge coupling relation is
$g^2_a=\frac{5}{6}\, g^2_b=\frac{35}{32} \,(\frac{5}{3}\,g^2_Y)= \frac{8\times 5^{3/4}\sqrt{7}}{27} \, \pi \,e^{\phi_4}$ \cite{Li:2019nvi}.}
	\label{tab:16}
	\begin{center}
		\begin{tabular}{|c||c|c||c|c|c|c|c|c|c|c|}
			\hline	 \rm{Model 16} &
			\multicolumn{10}{|c|}{${\rm U}(4)\times {\rm U}(2)_L\times {\rm U}(2)_R\times {\rm USp}(2)^2$}\\
			\hline \hline \rm{stack} & $N$ & $(n^1, l^1)\times (n^2, l^2)\times (n^3, l^3)$ & $n_{\yng(2)}$ & $n_{\yng(1,1)_{}}$ & $b$ & $b'$ & $c$ & $c'$& 2 & 3  \\
			\hline
			$a$&  8& $(1,-1)\times (1,1)\times (1,-1)$ & 0 & -4  & 3 & 0 & -3 & 0 & -1 & 1  \\
			$b$&  4& $(-2,5)\times (-1,0)\times (1,1)$ & 3 & -3  & - & - & 0 & -1 & -5 & 0   \\
			$c$&  4& $(-1,-2)\times (0,-1)\times (-1,-1)$ & -1 & 1  & - & - & - & - & 0 & -1  \\
			\hline
			2&   2& $(1,0)\times (0,-1)\times (0,2)$ & \multicolumn{8}{c|}{$\chi_1=\frac{\sqrt{5}}{5},\; \chi_2=\frac{7\sqrt{5}}{5},\; \chi_3=\sqrt{5}$}\\
			3&   2& $(0,-1)\times (1,0)\times (0,2)$ & \multicolumn{8}{c|}{$\beta^g_2=1,\; \beta^g_3=-3$}\\
			\hline
		\end{tabular}
	\end{center}
\end{table}

\begin{table}[h]
\footnotesize\renewcommand{\arraystretch}{1.3}
\caption{The chiral and vector-like superfields, and their quantum numbers under the gauge symmetry ${\rm SU}(4)_C\times {\rm SU}(2)_L\times {\rm SU}(2)_R \times {\rm USp}(2)_1 \times {\rm USp}(2)_2$.}
\label{tab:MSSM}
\begin{center}
\begin{tabular}{|c||c||c|c|c||c|c|c|}\hline
 & Quantum Number
& $Q_4$ & $Q_{2L}$ & $Q_{2R}$  & Field \\
\hline\hline
$ab$ & $3 \times (4,\overline{2},1,1,1)$ & 1 & -1 & 0  & $F_L(Q_L, L_L)$\\
$ac$ & $3\times (\overline{4},1,2,1,1)$ & -1 & 0 & $1$   & $F_R(Q_R, L_R)$\\
$bc'$ & $1 \times (1,\overline{2},\overline{2},1,1)$ & 0 & -1 & -1   & $H'$\\
$a2$ & $1\times (\overline{4},1,1,2,1)$ & -1 & 0 & 0  & $X_{a2}$ \\
$a3$ & $1\times (4,1,1,1,\overline{2})$ & 1 & 0 & 0   & $X_{a3}$ \\
$b2$ & $5\times(1,\overline{2},1,2,1)$ & 0 & -1 & 0    & $X_{b2}$ \\
$c3$ & $1\times(1,1,\overline{2},1,2)$ & 0 & 0 & -1    & $X_{c3}$ \\
$a_{\overline{\yng(1,1)}}$ & $4\times(\overline{6},1,1,1,1)$ & -2 & 0 & 0   & $S_C^i$ \\
$b_{\yng(2)}$ & $3\times(1,3,1,1,1)$ & 0 & 2 & 0   &  $T_L^i$ \\
$b_{\overline{\yng(1,1)}}$ & $3\times(1,\overline{1},1,1,1)$ & 0 & -2 & 0   & $S_L^i$ \\
$c_{\overline{\yng(2)}}$ & $1\times(1,1,\overline{3},1,1)$ & 0 & 0 & -2   & $T_R^i$  \\
$c_{\yng(1,1)_{}}$ & $1\times(1,1,1,1,1)$ & 0 & 0 & 2   & $S_R^i$ \\
\hline\hline
$bc$ & $9 \times (1,\overline{2},2,1,1)$ & 0 & -1 & 1   & $H_u^i$, $H_d^i$\\
     & $9 \times (1,2,\overline{2},1,1)$ & 0 & 1 & -1   & \\
\hline
\end{tabular}
\end{center}
\end{table}

Here we study the phenomenology of the newly found models where one of the wrapping numbers is 5. For concreteness, we choose Model 16 from ref.~\cite{Li:2019nvi}, which is T-dual to Model 18, see table~\ref{tab:16}. The model exhibits approximate gauge coupling unification with three-generations of chiral fermions together besides the hidden sector with gauge group ${\rm USp}(2)^2$. The detailed spectrum of chiral and vectorlike superfields of the model with their respective quantum numbers under the gauge symmetry ${\rm U}(4)_C\times {\rm U}(2)_L\times {\rm U}(2)_R \times {\rm USp}(2)_1 \times {\rm USp}(2)_2$ is listed in table~\ref{tab:MSSM}.

Placing the $a'$, $b$ and $c$ stacks of D6-branes on the top of each other on the third 2-torus results in additional vector-like particles from $\mathcal{N} = 2$ subsectors \cite{Cvetic:2004ui}. The anomalies from three global ${\rm U}(1)$s of ${\rm U}(4)_C$, ${\rm U}(2)_L$ and ${\rm U}(2)_R$ are cancelled by the Green-Schwarz mechanism, and the gauge fields of these ${\rm U}(1)$s obtain masses via the linear $B\wedge F$ couplings. Thus, the effective gauge symmetry is ${\rm SU}(4)_C\times {\rm SU}(2)_L\times {\rm SU}(2)_R$.

\section{Supersymmetry breaking and $\mathcal{N}=1$ Effective theory}\label{sec:EFT}

\begin{figure}[h]
\centering
\includegraphics[width=\textwidth]{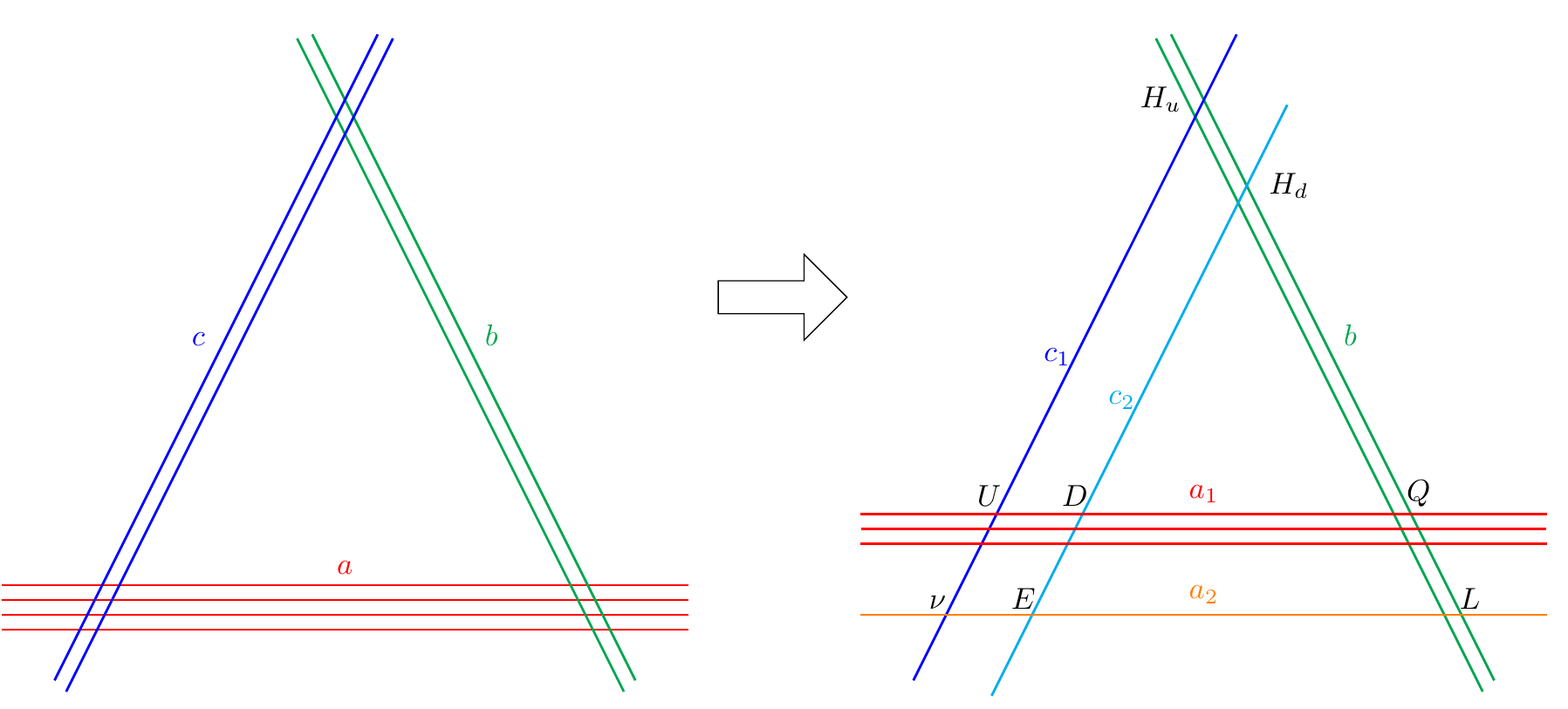}
\caption{Pati-Salam gauge group ${\rm SU}(4)\times {\rm SU}(2)_L\times {\rm SU}(2)_R$ is broken down to the standard model gauge group ${\rm SU}(3)_C\times {\rm U}(2)_L\times {\rm U}(1)_{I3R}\times {\rm U}(1)_{B-L}$ via the process of brane splitting that corresponds to assigning VEVs to the adjoint scalars, which arise as open-string moduli associated with the positions of stacks $a$ and $c$ in the internal space.} \label{brnsplit}
\end{figure}

Pati-Salam gauge group ${\rm SU}(4)\times {\rm SU}(2)_L\times {\rm SU}(2)_R$ is higgsed down to the standard model gauge group ${\rm SU}(3)_C\times {\rm U}(2)_L\times {\rm U}(1)_{I3R}\times {\rm U}(1)_{B-L}$ by assigning vacuum expectation values to the adjoint scalars which arise as open-string moduli associated to the stacks $a$ and $c$, see figure~\ref{brnsplit},
\begin{eqnarray}
\textcolor{red}{a} &\rightarrow & \textcolor{red}{a_1} + \textcolor{orange}{a_2} , \nonumber \\
\textcolor{blue}{c} &\rightarrow & \textcolor{blue}{c_1} + \textcolor{cyan}{c_2} .
\end{eqnarray}
Moreover, the ${\rm U}(1)_{I_{3R}}\times {\rm U}(1)_{B-L}$ gauge symmetry may
be broken to ${\rm U}(1)_Y$ by giving vacuum expectation values (VEVs)
to the vector-like particles with the quantum numbers $({\bf { 1},
1, 1/2, -1})$ and $({\bf { 1}, 1, -1/2, 1})$ under the
${\rm SU}(3)_C\times {\rm SU}(2)_L\times {\rm U}(1)_{I_{3R}} \times {\rm U}(1)_{B-L} $
gauge symmetry from $a_2 c_1'$
intersections~\cite{Cvetic:2004ui,Chen:2006gd}.

This brane-splitting results in standard model quarks and leptons as \cite{Cvetic:2004nk},
\begin{eqnarray}
F_L(Q_L, L_L)  &\rightarrow &  Q_L + L , \nonumber \\
F_R(Q_R, L_R)  &\rightarrow &  U_R + D_R + E_R + N.
\end{eqnarray}
Three-point Yukawa couplings for the quarks and the charged leptons can be read from the following superpotential,
\begin{equation}\label{eq:WY}
W_Y = Y^U_{ijk} Q_L^i U_R^j H_U^k + Y^D_{ijk} Q_L^i D_R^j H_D^k + Y^L_{ijk} L^i E^j H_D^k.
\end{equation}

The additional exotic particles must be made superheavy to ensure gauge coupling unification at the GUT scale.
Similar to Refs.~\cite{Cvetic:2007ku, Chen:2007zu} we can decouple  the additional exotic particles except
the four chiral multiplets under ${\rm SU}(4)_C$ anti-symmetric representation. And these four chiral multiplets can
be decoupled via instanton effects in principle \cite{Blumenhagen:2006xt, Haack:2006cy, Florea:2006si}, and
we will present the detailed discussions elsewhere.

We now turn our attention toward the four dimensional low energy effective field theory. $\mathcal{N}=1$ supergravity action is encoded by three functions viz. the gauge kinetic function $f_{x}$, the K\"{a}hler potential $K$ and the superpotential $W$ \cite{Cremmer:1982en}. Each of these functions in turn depend on dilaton $S$, complex $T$, and K\"{a}hler $U$ moduli.

The complex structure moduli $U$ can be obtained from the supersymmetry conditions as,
\begin{align}\label{U-moduli}
U^i & = \frac{i R_2^i}{R_1^i+\frac{\beta_i}{2} iR_2^i} = \frac{i \chi^i}{1+\frac{\beta_i}{2} i\chi^i}, \qquad \because \chi^i \equiv \frac{R_2^i}{R_1^i}.
\end{align}
These upper case moduli in string theory basis can be transformed in to lower case {$s$, $t$, $u$} moduli in field theory basis as \cite{Lust:2004cx},
\begin{align}
\mathrm{Re}\,(s) &= \frac{e^{-{\phi}_4}}{2\pi}\,\left(\frac{\sqrt{\mathrm{Im}\,U^{1}\, \mathrm{Im}\,U^{2}\,\mathrm{Im}\,U^3}}{|U^1U^2U^3|}\right) , \nonumber \\
\mathrm{Re}\,(u^j) &= \frac{e^{-{\phi}_4}}{2\pi}\left(\sqrt{\frac{\mathrm{Im}\,U^{j}}{\mathrm{Im}\,U^{k}\,\mathrm{Im}\,U^l}}\right)\; \left|\frac{U^k\,U^l}{U^j}\right|, \quad (j,k,l)=(\overline{1,2,3}), \nonumber \\
\mathrm{Re}(t^j) &= \frac{i\alpha'}{T^j} , \label{eq:moduli}
\end{align}
where $j$ denotes the $j^{\rm th}$ two-torus, and $\phi_4$ is the four dimensional dilaton which is related to the supergravity moduli as \footnote{There was a typo in \cite{Kane:2004hm} in the paragraph after equation (18) where $\phi_4$ is related to the supergravity moduli.}
\begin{equation}
2\pi e^{\phi_4}=\Big(\mathrm{Re}(s)\,\mathrm{Re}(u_1)\,\mathrm{Re}(u_2)\,\mathrm{Re}(u_3)\Big)^{-1/4}.
\end{equation}
Inverting the above formulas we can solve for $U$ moduli in string theory basis in terms of $s$ and $u$ as,
\begin{equation}
\frac{|U^j|^2}{\mathrm{Im}\,(U^j)} = \sqrt{\frac{\mathrm{Re}\,(u^k)\,\mathrm{Re}\,(u^l)}{\mathrm{Re}\,(u^j)\mathrm{Re}\,(s)}}, \quad  (j,k,l)=(\overline{1,2,3}) . \label{eq:b}
\end{equation}
The holomorphic gauge kinetic function for any D6-brane stack $x$ wrapping a calibrated 3-cycle is given as \cite{Blumenhagen:2006ci},
\begin{equation}
f_x = \frac{1}{2\pi \ell_s^3}\left[e^{-\phi}\int_{\Pi_x} \mbox{Re}(e^{-i\theta_x}\Omega_3)-i\int_{\Pi_x}C_3\right],
\end{equation}
where the integral involving 3-form $\Omega_3$ gives,
\begin{equation}
\int_{\Pi_x}\Omega_3 = \frac{1}{4}\prod_{i=1}^3(n_x^iR_1^i + 2^{-\beta_i}il_x^iR_2^i).
\end{equation}
It can then be shown that,
\begin{eqnarray}
f_x &=&
\frac{1}{4\kappa_x}(n_x^1\,n_x^2\,n_x^3\,s-\frac{n_x^1\,l_x^2\,l_x^3\,u^1}{2^{(\beta_2+\beta_3)}}-\frac{l_x^1\,n_x^2\,l_x^3\,u^2}{2^{(\beta_1+\beta_3)}}-
\frac{l_x^1\,l_x^2\,n_x^3\,u^3}{2^{(\beta_1+\beta_2)}}),
\label{kingauagefun}
\end{eqnarray}
where the factor ${\kappa}_x$ is related to the difference between the gauge couplings for ${\rm U}(N_x)$ and $Sp(2N_x),\,SO(2N_x)$. ${\kappa}_x =1$ for ${\rm U}(N_x)$ and ${\kappa}_x =2$ for $Sp(2N_x)$ or $SO(2N_x)$ \cite{Klebanov:2003my}. Since, the standard model hypercharge ${\rm U}(1)_Y$ is a linear combination of several ${\rm U}(1)$s,
\begin{equation}
Q_Y=\frac{1}{6}Q_{a_1}+\frac{1}{2}Q_{a_2}-\frac{1}{2}Q_{c_1}-\frac{1}{2}Q_{c_2}.
\end{equation}
Therefore, the holomorphic gauge kinetic function for the hypercharge is also taken as a linear combination of the kinetic gauge functions from all of the stacks as \cite{Blumenhagen:2003jy, Ibanez:2001nd},
\begin{equation}\label{fY}
f_Y=\frac{1}{6}f_{a_1}+\frac{1}{2}f_{a_2}+\frac{1}{2}f_{c_1}+\frac{1}{2}f_{c_2}.
\end{equation}
The K\"{a}hler potential to the second order for the moduli $M$ and open string matter fields $C_i, C_\theta$ is given by :
\begin{align}
K(M,\bar{M},C,\bar{C}) &= \hat{K}(M,\bar{M}) +  \sum_{\mathrm{untwisted}\,i,j} \tilde{K}_{C_{i}\bar{C}_j}(M,\bar{M}) C_i\bar{C}_j \nonumber\\
& \quad  + \sum_{\mathrm{twisted}, \, \theta} \tilde{K}_{C_{\theta}\bar{C}_{\theta}}(M,\bar{M}) C_{\theta}\bar{C}_{\theta} ,
\end{align}
where $C_i$ correspond to the D-brane positions and the Wilson lines moduli arising from strings having both ends on the same stack while $C_{\theta}$ correspond to strings stretching between different stacks comprising $1/4$ BPS branes. The untwisted moduli fields $C_{i},\bar{C}_j$ are not present in MSSM and must become heavy via higher dimensional operators\footnote{D-branes wrapping rigid cycles can freeze such open string moduli \cite{Blumenhagen:2005tn}, however such rigid cycles without discrete torsion are not present in $\mathbf{T}^6/(\mathbb{Z}_2 \times \mathbb{Z}_2)$.}.

Let us determine the K\"{a}hler metric $\tilde{K}_{C_{\theta}\bar{C}_{\theta}}(M,\bar{M})$ for the twisted moduli. We denote the K\"{a}hler potential arising from strings stretching between stacks $x$ and $y$ as $\tilde{K}_{xy}$ and $\theta^j_{xy}\equiv \theta^j_y-\theta^j_x$ denotes the angle between the cycles wrapped by the branes $x$ and $y$ on the $j^{\rm th}$ two-torus with the constraint $\sum_j\theta^j_{xy}=0$. Following \cite{Font:2004cx,Cvetic:2003ch,Lust:2004cx}, we find two cases for the K\"{a}hler metric in type IIA theory:
\begin{itemize}
\item $\theta^j_{xy}<0$, $\theta^k_{xy}>0$, $\theta^l_{xy}>0$
\begin{eqnarray}
\tilde{K}_{xy} &=& e^{\phi_4} e^{\gamma_E (2-\sum_{j = 1}^3
\theta^j_{xy}) }
\sqrt{\frac{\Gamma(\theta^j_{xy})}{\Gamma(1+\theta^j_{xy})}}
\sqrt{\frac{\Gamma(1-\theta^k_{xy})}{\Gamma(\theta^k_{xy})}}
\sqrt{\frac{\Gamma(1-\theta^l_{xy})}{\Gamma(\theta^l_{xy})}}
\nonumber \\ && (t^j + \bar{t}^j)^{\theta^j_{xy}} (t^k +
\bar{t}^k)^{-1+\theta^k_{xy}} (t^l +
\bar{t}^l)^{-1+\theta^l_{xy}}. \label{eq:kahler1}
\end{eqnarray}
\item $\theta^j_{xy}<0$, $\theta^k_{xy}<0$, $\theta^l_{xy}>0$
\begin{eqnarray}
\tilde{K}_{xy} &=& e^{\phi_4} e^{\gamma_E (2+\sum_{j = 1}^3
\theta^j_{xy}) }
\sqrt{\frac{\Gamma(1+\theta^j_{xy})}{\Gamma(-\theta^j_{xy})}}
\sqrt{\frac{\Gamma(1+\theta^k_{xy})}{\Gamma(-\theta^k_{xy})}}
\sqrt{\frac{\Gamma(\theta^l_{xy})}{\Gamma(1-\theta^l_{xy})}}
\nonumber \\ && (t^j + \bar{t}^j)^{-1-\theta^j_{xy}} (t^k +
\bar{t}^k)^{-1-\theta^k_{xy}} (t^l + \bar{t}^l)^{-\theta^l_{xy}}. \label{eq:kahler2}
\end{eqnarray}
\end{itemize}
The K\"{a}hler metric for the branes parallel to at least one torus which give rise to non-chiral matter in bifundamental representations (1/2 BPS scalar) like the Higgs doublet is,
\begin{equation}
\hat{K}_\mathrm{higgs} =\left[(s+\bar{s})(t^1+\bar{t}^1)(t^2+\bar{t}^2)(u^3+\bar{u}^3)\right]^{-1/2}.  \label{nonChiralK}
\end{equation}
The superpotential is given as,
\begin{equation}
W = \hat{W} + \frac{1}{2} {\mu}_{\alpha \beta}(M)\, C^{\alpha}\,C^{\beta}+ \frac{1}{6}\,Y_{\alpha\beta\gamma}(M)\,C^{\alpha}\,C^{\beta}\,C^{\gamma}+... \label{eq:W}
\end{equation}
and the minimum of the tree-level F-term supergravity scalar potential is given by\footnote{In our analysis we assume that D-terms do not affect the soft terms \cite{Kawamura:1996ex, Komargodski:2009pc}.}
\begin{align}
V(M,\bar{M}) &= e^G(G_M K^{MN} G_N -3) \nonumber\\
             &= (F^M K_{MN} F^N-3 e^G),
\end{align}
where $G_M=\partial_M G$, $K_{MN}=\partial_M \partial_N K$, $K^{MN}$ is the inverse K\"{a}hler metric, and the auxiliary fields $F^M$ are,
\begin{equation}
F^M=e^{G/2} K^{MN}G_N. \label{aux}
\end{equation}
Thus supersymmetry is broken via F-terms from some of the hidden sector fields $M$ acquiring VEVs, thereby generating soft terms in the observable sector \cite{Font:2004cx, Kane:2004hm, Chen:2007zu}. Gravitino gets massive by absorbing Goldstino via the superhiggs mechanism.
\begin{equation}
m_{3/2}=e^{G/2}.
\end{equation}
The \emph{normalized} soft parameters viz. the gaugino mass, squared scalar mass and trilinear parameters are given by \cite{Brignole:1997dp},
\begin{align}
M_x &= \frac{1}{2\,\mathrm{Re}\,f_x}\, (F^M\,\partial_M\,f_x), \nonumber \\
m_{xy}^2 &= (m_{3/2}^2 + V_0) - \sum_{M,N}\, \bar{F}^{\bar{M}}F^N\,{\partial}_{\bar{M}}\,{\partial}_{N}\,\log({\tilde{K}}_{xy}), \nonumber \\
A_{xyz} &= F^M[\hat{K}_M+{\partial}_M\,\log(Y_{xyz})-{\partial}_{M}\,\log(\tilde{K}_{xy}\tilde{K}_{yz}\tilde{K}_{zx})], \label{softterms}
\end{align}
where $\hat{K}_M$ is the K\"{a}hler metric for branes parallel to at least one torus and $F^M$ denotes auxiliary fields.

Although it appears that soft terms may depend on the Yukawa couplings via the superpotential, however these are \emph{not} the physical Yukawa couplings which exponentially depend on the worldsheet area as discussed in section~\ref{sec:Yukawa}. Both are related by the following relation,
\begin{equation}
Y^{\mathrm{phys}}_{xyz} = Y_{xyz}\, \frac{\hat{W}^*}{|\hat{W}|}\,e^{\hat{K}/2}\, (\tilde{K}_{x}\tilde{K}_{y}\tilde{K}_{z})^{-1/2}.
\end{equation}

To calculate the soft terms from supersymmetry breaking we ignore the cosmological constant $V_0$ and introduce the following VEVs for the auxiliary fields \eqref{aux} for the $s$, $t$ and $u$ moduli \cite{Brignole:1993dj},
\begin{align}\label{eq:aux}
F^s &= 2\sqrt{3}C m_{3/2} \mathrm{Re}(s) \Theta_s e^{-i\gamma_s}, \nonumber \\
F^{\{u,t\}^i} &= 2\sqrt{3}C m_{3/2}\left( \mathrm{Re}  ({u}^i) \Theta_i^u e^{-i\gamma^u_i}+  \mathrm{Re} ({t}^i) \Theta_i^t e^{-i\gamma_i^t}\right),
\end{align}
Here, the factors $\gamma_s$ and $\gamma_i$ denote the CP violating phases of the moduli. The constant $C$ is given by the gravitino-mass $m^2_{3/2}$ and the cosmological constant $V_0$ as $C^2 = 1+ \frac{V_0}{3 m^2_{3/2}}$. $\Theta_s$ and $\Theta^{t,u}_i$ are the goldstino angles which determine the degree to which supersymmetry breaking is being dominated by any of the dilaton $s$, complex structure ($u^i$) and K\"ahler ($t^i$) moduli constrained by the relation,
\begin{align}
\sum_{i=1}^3 (|\Theta_i^u|^2 + |\Theta_i^t|^2) + |\Theta_s|^2 =1.
\end{align}
Unlike the $s$- or $u$-moduli dominant supersymmetry breaking, the case of $t$-moduli dominant susy breaking depends on the physical Yukawa couplings via the area of the triangles and thus we shall only concentrate on the following two scenarios:
\begin{enumerate}
    \item The $u$-moduli dominated supersymmetry breaking with the goldstino angle $\Theta_s$ set to zero, such that $F^s=F^{t^i}=0$.
    \item The $u$ and $s$-moduli supersymmetry breaking with $F^s \neq 0$ .
\end{enumerate}
The cosmological constant, $V_0$ is taken to be zero in all cases.

\subsection{Supersymmetry breaking with $u$-moduli dominance}
In the $u$-moduli dominant susy breaking $\Theta_s = 0$ and the auxiliary fields \eqref{eq:aux} become,
\begin{equation}
F^{u^i} = \sqrt{3}m_{3/2}(u^i + \bar{u}^i)\Theta_i e^{-i\gamma_i},\quad i=1,2,3. \label{auxfields}
\end{equation}
To calculate the soft terms, we need to know the derivatives of the K\"{a}hler potential with respect to $u$. Defining $\tilde{K}_{xy}\equiv e^{{\phi}_4}\,\tilde{K}^0_{xy}$ and using (\ref{eq:kahler1}) and \eqref{eq:kahler2}, we compute the derivatives with respect to $u^i$ as,
\begin{eqnarray}
\frac{\partial
\log{\tilde{K}_{xy}}}{\partial u^i}&=& \sum_{j=1}^3\frac {\partial
\log{\tilde{K}^0_{xy}}}{\partial\theta^j_{xy}}
\frac{\partial\theta^j_{xy}}{\partial u^i} - \frac{1}{4(u^i+\bar{u}^i)}, \\
\frac{\partial^2 \log{\tilde{K}_{xy}}}{\partial u^i\partial\bar ,u^j}&=& \sum_{k=1}^3\left(\frac{\partial
\log{\tilde{K}^0_{xy}}}{\partial\theta^k_{xy}}
\frac{\partial^2\theta^k_{xy}}{\partial u^i\partial\bar u^j}+
\frac{\partial^2\log \tilde{K}^0_{xy}}{\partial (\theta^k_{xy})^2}
\frac{\partial\theta^k_{xy}}{\partial u^i}
\frac{\partial\theta^k_{xy}}{\partial \bar
u^j}+\frac{{\delta}_{ij}}{4\,(u^i+\bar{u}^i)^2}\right). \nonumber
\end{eqnarray}
From the K\"{a}hler potential in \eqref{eq:kahler2}, we have
\begin{align}
\Psi(\theta^j_{xy}) &\equiv \frac {\partial
\log{\tilde{K}^0_{xy}}}{\partial\theta^j_{xy}}=
\gamma_E\!+\!\frac{1}{2}\frac{d}{d{\theta}^j_{xy}}\log{\Gamma(1-\theta^j_{xy})}-
\frac{1}{2}\frac{d}{d{\theta}^j_{xy}}\log{\Gamma(\theta^j_{xy})}-\log(t^j+\bar t^j), \label{eq:psi}\\
\Psi'(\theta^j_{xy}) &\equiv \frac{\partial^2\log
\tilde{K}^0_{xy}} {\partial(\theta^j_{xy})^2}=
\frac{d\Psi(\theta^j_{xy})}{d \theta^j_{xy}}.  \label{eq:dpsi}
\end{align}
The angles $\theta^j_{xy}\equiv \theta^j_y-\theta^j_x$ are related to the $u$ moduli as,
\begin{equation}\label{Eq:angle}
\tan(\pi\theta^j_x)=\frac{2^{-\beta_j}l^j_x}{n^j_x}
\sqrt{\frac{\mathrm{Re}u^k\,\mathrm{Re}u^l}{\mathrm{Re}u^j\,\mathrm{Re}s}} \quad\mathrm{where~}
(j,k,l)=(\overline{1,2,3}).
\end{equation}
And the derivative of the angles are defined as,
\begin{equation}
{\theta}^{j,k}_{xy} \equiv (u^k+\bar u^k)\,\frac{\partial
\theta^j_{xy}}{\partial u^k}= \left\{\begin{array}{l}
 \left[-\frac{1}{4\pi}
 \sin(2\pi\theta^j)
 \right]^x_y \quad \mathrm{when~}j=k  \quad \\
 \left[\frac{1}{4\pi}
\sin(2\pi\theta^j)
 \right]^x_y \quad \mathrm{when~}j\neq k
\end{array}\right.\label{eq:dthdu}
\end{equation}
where $[f(\theta^j)]^x_y=f(\theta^j_x)-f(\theta^j_y)$.
And the second order derivatives become,
\begin{equation}
{\theta}^{j,k\bar{l}}_{xy} \equiv (u^k+\bar
u^k)(u^l+\bar u^l)\,\frac{\partial^2 \theta^j_{xy}}{\partial
u^k\partial\bar u^l}= \left\{\begin{array}{l}
\frac{1}{16\pi}
  \left[ \sin(4\pi\theta^j)+4\sin(2\pi\theta^j)
 \right]^x_y \quad
   \mathrm{when~}j=k=l  \\
 \frac{1}{16\pi}  \left[
 \sin(4\pi\theta^j)-4\sin(2\pi\theta^j)
 \right]^x_y \quad
   \mathrm{when~}j\neq k=l \\
 -\frac{1}{16\pi}\left[
 \sin(4\pi\theta^j)
 \right]^x_y \quad
   \mathrm{when~}j=k\neq l \mathrm{~or~} j=l\neq k \\
 \frac{1}{16\pi}\left[
\sin(4\pi\theta^j)
 \right]^x_y \quad
   \mathrm{when~}j\neq k\neq l\neq j
\end{array}\right.\label{eq:d2thdu2}
\end{equation}
We can now substitute the parametrizations \eqref{auxfields}-\eqref{eq:dpsi} in the general expressions \eqref{softterms} to calculate the soft terms:
\begin{itemize}
\item Gaugino mass parameters:\footnote{There was a typo in \cite{Chen:2007zu} in equation (29), the factor $2^{-(\beta_k +\beta_l)}$ was missed.}
\begin{eqnarray}
M_x=\frac{-\sqrt{3}m_{3/2}}{4\mathrm{Re}f_x} \sum_{j=1}^3 \left( \mathrm{Re}u^j\,\Theta_j\, e^{-i\gamma_j}\,2^{-(\beta_k +\beta_l)}n^j_x l^k_x l^l_x\right), \nonumber\\
(j,k,l)=(\overline{1,2,3}). \label{eq:idb:gaugino}
\end{eqnarray}
Bino mass parameter is then related to the linear combination of the gaugino masses for each stack as,
\begin{equation}\label{Bino-mass}
M_Y = \frac{1}{f_Y}\sum_x c_x f_x M_x ,
\end{equation}
where the coefficients $c_x$ correspond to the linear combination of ${\rm U}(1)$ factors which define the hypercharge, ${\rm U}(1)_Y = \sum c_x {\rm U}(1)_x$, cf. \eqref{fY}.
\item Trilinear parameters:
\begin{align}
A_{xyz}&=-\sqrt{3}m_{3/2}\sum_{j=1}^3 \left[\Theta_je^{-i\gamma_j}\left(\frac{1}{2}+\sum_{k=1}^3 \theta_{xy}^{k,j}\Psi(\theta^k_{xy})+\sum_{k=1}^3
 \theta_{zx}^{k,j}\Psi(\theta^k_{zx})\right)\right]  \nonumber \\
&\quad +\frac{\sqrt{3}}{2}m_{3/2}{\Theta}_{3}e^{-i{\gamma}_1} ,
\end{align}
where $x$, $y$, and $z$ label those stacks of branes whose intersections define the corresponding fields present in the trilinear coupling. Since the differences of the angles may be negative $\theta_{xy} = \theta_y - \theta_x$, it is useful to define the sign parameter,
\begin{equation}
\eta_{xy} = \prod_i (-1)^{1-H(\theta_{xy}^{i})},\quad H(x)= \begin{cases}
      0, & x < 0 \\
      1, & x\geq 0
    \end{cases}
\end{equation}
where the value $\eta_{xy} = -1$ indicates that only one of the angle differences is negative while $\eta_{xy} = +1$ indicates that two of the angle differences are negative.
\item Squarks and sleptons mass-squared (1/4 BPS scalars):
\begin{eqnarray}
m^2_{xy}= m_{3/2}^2\left[1-3\sum_{m,n=1}^3 \Theta_m\Theta_n e^{-i(\gamma_m-\gamma_n)}\left( \frac{{\delta}_{mn}}{4}+ \sum_{j=1}^3 \left(\theta^{j,m\bar
n}_{xy}\Psi(\theta^j_{xy})+ \theta^{j,m}_{xy}\theta^{j,\bar n}_{xy}\Psi'(\theta^j_{xy})\right)\right)\right]. \nonumber\\
\end{eqnarray}
Here, the functions $\Psi(\theta_{xy})=\frac{\partial \log (e^{-\phi_4}\tilde{K}_{xy})}{\partial \theta_{xy}}$ in the case of $\eta_{xy}=-1$ are
\begin{eqnarray}\label{eqn:Psi1}
\mathrm{if} \ \theta_{xy} < 0&:& \\
\Psi(\theta^j_{xy})&=&
-\gamma_E+\frac{1}{2}\frac{d}{d{\theta}^j_{xy}}\,\log{\Gamma(-\theta^j_{xy})}-
\frac{1}{2}\frac{d}{d{\theta}^j_{xy}}\,\log{\Gamma(1+\theta^j_{xy})}+\log(t^j+\bar t^j)\nonumber\\
\mathrm{if} \ \theta_{xy} > 0&:& \nonumber\\
\Psi(\theta^j_{xy})&=&
-\gamma_E+\frac{1}{2}\frac{d}{d{\theta}^j_{xy}}\,\log{\Gamma(1-\theta^j_{xy})}-
\frac{1}{2}\frac{d}{d{\theta}^j_{xy}}\,\log{\Gamma(\theta^j_{xy})}+\log(t^j+\bar t^j),\nonumber
\end{eqnarray}
and in the case of $\eta_{xy}=+1$ are
\begin{eqnarray}\label{eqn:Psi2}
\mathrm{if} \ \theta_{xy} < 0&:&  \\
\Psi(\theta^j_{xy})&=&
\gamma_E+\frac{1}{2}\frac{d}{d{\theta}^j_{xy}}\,\log{\Gamma(1+\theta^j_{xy})}-
\frac{1}{2}\frac{d}{d{\theta}^j_{xy}}\,\log{\Gamma(-\theta^j_{xy})}-\log(t^j+\bar t^j)  \nonumber\\
\mathrm{if} \ \theta_{xy} > 0&:&  \nonumber\\
\Psi(\theta^j_{xy})&=&
\gamma_E+\frac{1}{2}\frac{d}{d{\theta}^j_{xy}}\,\log{\Gamma(\theta^j_{xy})}-
\frac{1}{2}\frac{d}{d{\theta}^j_{xy}}\,\log{\Gamma(1-\theta^j_{xy})}-\log(t^j+\bar t^j) ,\nonumber
\end{eqnarray}
and $\Psi'(\theta_{xy})$ is just the derivative $\Psi'(\theta^j_{xy}) =\frac{d\Psi(\theta^j_{xy})}{d \theta^j_{xy}}$.

\item Higgs (1/2 BPS scalar) mass-squared is computed using the K\"{a}hler metric \eqref{nonChiralK} as:
\begin{equation}\label{Higgs-mass}
m^2_H = m^2_{3/2}\left(1-\frac{3}{2}\left|\Theta_3\right|^2\right).
\end{equation}
\end{itemize}

\subsection{Supersymmetry breaking via $u$-moduli and dilaton $s$}
Now we also include a non-zero VEV for the dilaton $s$ in the auxiliary fields \eqref{eq:aux} to get,
\begin{equation}\label{auxfields_su}
F^{s,u^i} = \sqrt{3}m_{3/2}[(s + \bar{s})\Theta_s e^{-i\gamma_s} + (u^i + \bar{u}^i)\Theta_i e^{-i\gamma_i}].
\end{equation}
Substituting above parametrization \eqref{auxfields_su} and the expressions \eqref{auxfields}-\eqref{eq:dpsi} in the general formulas \eqref{softterms}, the soft parameters are found as follows:
\begin{itemize}
\item Gaugino mass parameters:
\begin{eqnarray}\label{gaugino-masses}
M_x=\frac{-\sqrt{3}m_{3/2}}{4\mathrm{Re} f_x}\Bigg[\sum_{j=1}^3
\mathrm{Re} (u^j)\,\Theta_j\,
e^{-i\gamma_j}\, 2^{-(\beta_k +\beta_l)}n^j_x l^k_x l^l_x  +\Theta_s\mathrm{Re}(s) e^{-i\gamma_0}n_x^1\,n_x^2\,n_x^3\, \Bigg], \nonumber\\
(j,k,l)=(\overline{1,2,3}).\qquad\qquad\qquad
\end{eqnarray}
and the Bino mass parameter is similarly defined as \eqref{Bino-mass}.
\item Trilinear parameters:
\begin{align}\label{tri-coupling}
A_{xyz}&=-\sqrt{3}m_{3/2}\sum_{j=1}^4 \left[
\Theta_je^{-i\gamma_j}\left(\frac{1}{2}+\sum_{k=1}^3
 \theta_{xy}^{k,j}\Psi(\theta^k_{xy})+\sum_{k=1}^3
 \theta_{zx}^{k,j}\Psi(\theta^k_{zx})
\right)\right] \nonumber \\
&\quad +\frac{\sqrt{3}}{2}m_{3/2}\left({\Theta}_{3}e^{-i{\gamma}_3} +
\Theta_s e^{-i{\gamma}_s}\right),
\end{align}
where $j=4$ corresponds to $\Theta_s$.
\item Squarks and sleptons mass-squared (1/4 BPS scalars):
\begin{align}\label{slepton-mass}
m^2_{xy}&= m_{3/2}^2\left[1-3\sum_{m,n=1}^4
\Theta_m\Theta_ne^{-i(\gamma_m-\gamma_n)}\left(
\frac{{\delta}_{mn}}{4}+ \sum_{j=1}^3 \left(\theta^{j,m\bar
n}_{xy}\Psi(\theta^j_{xy})+
 \theta^{j,m}_{xy}\theta^{j,\bar n}_{xy}\Psi'(\theta^j_{xy})\right)\right)
\right],\nonumber\\
\end{align}
where now $\Theta_4 \equiv \Theta_s$ is also included in the sum while the functions $\Psi(\theta_{xy})$, $\Psi'(\theta_{xy})$ and the terms ${\theta}^{j,k}_{xy}$,
${\theta}^{j,k\bar{l}}_{xy}$ are defined similarly as before in equations \eqref{eq:psi}, \eqref{eq:dpsi}, \eqref{eq:dthdu} and \eqref{eq:d2thdu2}. While the terms associated with the dilaton $s$ are given as,
\begin{equation}
{\theta}^{j,s}_{xy} \equiv (s+\bar s)\,\frac{\partial
\theta^j_{xy}}{\partial s}=
 -\frac{1}{4\pi}\left[
 \sin(2\pi\theta^j)
 \right]^x_y , \label{eq:dthdus2}
\end{equation}
\begin{equation}
{\theta}^{j,k\bar s}_{xy} \equiv (u^k+\bar
u^k)(s+\bar s)\,\frac{\partial^2 \theta^j_{xy}}{\partial
u^k\partial\bar s}= \left\{\begin{array}{l}
  \frac{1}{16\pi}\left[\sin{4\pi\theta^j}\right]^x_y \quad \mathrm{when} \ j=k \quad \\
  -\frac{1}{16\pi}\left[\sin{4\pi\theta^j}\right]^x_y \quad \mathrm{when} \ j\neq k,
\end{array}\right.\label{eq:dth2duds}
\end{equation}
and
\begin{equation}
{\theta}^{j,s\bar s}_{xy} \equiv (s+\bar
s)(s+\bar s)\,\frac{\partial^2 \theta^j_{xy}}{\partial
s\partial\bar s}=
  \frac{1}{16\pi}\left[\sin{4\pi\theta^j} + 4\sin(2\pi\theta^j)\right]^x_y,\label{eq:dth2dss}
\end{equation}
where $k,l\neq s$.
\item Higgs mass-squared (1/2 BPS scalar):
\begin{equation}
m^2_H = m^2_{3/2}\left[1-\frac{3}{2}\left(\left|\Theta_3\right|^2+\left|\Theta_s\right|^2\right)\right].
\end{equation}
\end{itemize}

\subsection{Soft terms from susy breaking}
We now systematically compute the soft terms from susy breaking in the general case of $u$-moduli dominance with dilaton modulus $s$ turned on. The complex structure moduli $U$ from \eqref{U-moduli} are,
\begin{align}\label{eqn:U-moduli}
\{U_1,U_2,U_3\} &=\left\{\frac{i}{\sqrt{5}},\frac{7 i}{\sqrt{5}},\frac{2}{9} \left(5+2 i \sqrt{5}\right)\right\} ,
\end{align}
and the corresponding $u$-moduli and $s$-modulus in supergravity basis from \eqref{eq:moduli} are,
\begin{align}\label{s-u-moduli}
\{u_1,u_2,u_3\} & = \left\{\frac{\sqrt[4]{5} \sqrt{7} e^{-\phi_4}}{2 \pi },\frac{\sqrt[4]{5} e^{-\phi_4}}{2 \sqrt{7} \pi },\frac{\sqrt{7} e^{-\phi_4}}{2\ 5^{3/4} \pi }\right\}, \nonumber\\
s & = \frac{\sqrt[4]{5} e^{-\phi_4}}{2 \sqrt{7} \pi } .
\end{align}
Using \eqref{kingauagefun} and the values from the table~\ref{tab:16}, the gauge kinetic function becomes,
\begin{align}\label{fx}
\{f_a,f_b,f_c\} & = \left\{\frac{27 e^{-\phi_4}}{8\ 5^{3/4} \sqrt{7} \pi },\frac{9 \sqrt[4]{5} e^{-\phi_4}}{16 \sqrt{7} \pi },\frac{9 \sqrt{7} e^{-\phi_4}}{16\ 5^{3/4} \pi }\right\},
\end{align}
To calculate gaugino masses $M_{1,2,3}$ respectively for ${\rm U}(1)_Y$, ${\rm SU}(2)_L$ and ${\rm SU}(3)_c$ gauge groups, we first compute $M_{a,b,c}$ using \eqref{gaugino-masses} as,
\begin{align}
M_a &= \frac{m_{3/2}}{18 \sqrt{3}}\Big(35 \Theta_1 e^{i \gamma_1}-5 \Theta_2 e^{-i \gamma_2}+14 \Theta_3 e^{-i\gamma_3}-10 \Theta_4 e^{-i\gamma_4}\Big),\nonumber \\
M_b &= \frac{m_{3/2}}{3 \sqrt{3}} \left(5 e^{-i \gamma_2} \Theta_2-4 e^{-i \gamma_4} \Theta_4\right),\nonumber \\
M_c &= \frac{m_{3/2}}{3 \sqrt{3}} \left(5 e^{-i \gamma_1} \Theta_1+4 e^{-i \gamma_3} \Theta_3\right).
\end{align}
Bino mass parameter $M_Y$ \eqref{Bino-mass} is then computed as,
\begin{align}
M_Y &=\frac{m_{3/2}}{99 \sqrt{3}} \left(175 \Theta _1 e^{-i \gamma_1}-10 \Theta _2 e^{-i \gamma_2}+112 \Theta _3 e^{-i \gamma_3}-20 \Theta _4 e^{-i \gamma_4}\right).
\end{align}
Therefore, the gaugino masses for ${\rm U}(1)_Y$, ${\rm SU}(2)_L$ and ${\rm SU}(3)_c$ gauge groups are,
\begin{align}\label{Gauginos123}
M_1 &\equiv M_Y = \frac{m_{3/2}}{99 \sqrt{3}} \left(175 \Theta _1 e^{-i \gamma_1}-10 \Theta _2 e^{-i \gamma_2}+112 \Theta _3 e^{-i \gamma_3}-20 \Theta _4 e^{-i \gamma_4}\right), \nonumber\\
M_2 &\equiv M_b = \frac{m_{3/2}}{3 \sqrt{3}} \left(5 e^{-i \gamma_2} \Theta_2-4 e^{-i \gamma_4} \Theta_4\right),\nonumber \\
M_3 &\equiv M_a = \frac{m_{3/2}}{18 \sqrt{3}}\Big(35 \Theta_1 e^{i \gamma_1}-5 \Theta_2 e^{-i \gamma_2}+14 \Theta_3 e^{-i\gamma_3}-10 \Theta_4 e^{-i\gamma_4}\Big).
\end{align}
\begin{table}[t] \footnotesize
\renewcommand{\arraystretch}{1.0}
\caption{The angles (in multiples of $\pi$) with respect to the orientifold plane
made by the cycle wrapped by each stack of D-branes
on each of the three two-tori.}
\label{Angles} \centering
\begin{tabular}{|c|c|c|c|}\hline
 & $\pi\theta^1$ &  $\pi\theta^2$  & $\pi\theta^3$ \\
\hline
$a$ & $-\tan ^{-1}\left(\frac{1}{\sqrt{5}}\right) $ & $ \tan ^{-1}\left(\frac{7}{\sqrt{5}}\right)$ & $-\tan ^{-1}\left(\frac{\sqrt{5}}{2}\right) $ \\
\hline
$b$ &  $-\tan ^{-1}\left(\frac{\sqrt{5}}{2}\right)$ & $0$ & $\tan ^{-1}\left(\frac{\sqrt{5}}{2}\right) $ \\
\hline
$c$ &  $\tan ^{-1}\left(\frac{2}{\sqrt{5}}\right)$ & $-\frac{\pi}{2}$ & $\tan ^{-1}\left(\frac{\sqrt{5}}{2}\right)$ \\
\hline
\end{tabular}
\end{table}
\par Next, to compute the trilinear coupling and the sleptons mass-squared we require the angles, the differences of angles and their first and second order derivatives with respect to the moduli. In table~\ref{Angles} we show the angles made by the cycle wrapped by each stack D6 branes with respect to the orientifold plane,
\begin{align}\label{eqn:Angle}
\pi \theta_{x}^i &= \tan^{-1}\left(\frac{2^{-\beta_i}l_x^i\chi_i}{n_x^i}\right).
\end{align}
The differences of the angles, $\theta_{xy}^{i}\equiv \theta_{y}^{i} -\theta_{x}^{i} $ are,
\begin{align}\label{angle-diff}
\left(\begin{array}{ccc}
 \hskip-5em\{0.,0.,0.\} & \hskip-7em \{-0.13386,-0.401581,0.535441\} & \{0.36614,-0.901581,0.535441\} \\
 \{0.13386,0.401581,-0.535441\} & \{0.,0.,0.\} & \{0.5,-0.5,0.\} \\
 \{-0.36614,0.901581,-0.535441\} & \{-0.5,0.5,0.\} & \{0.,0.,0.\} \\
\end{array}\right)
\end{align}
\par To account for the negative angle differences it is convenient to define the sign function $\sigma_{xy}^{i}$, which is $-1$ only for negative angle difference and $+1$ otherwise,
\begin{align}\label{sigmaK}
\sigma_{xy}^{i} \equiv (-1)^{1-H(\theta_{xy}^{i})} & = \left(
\begin{array}{ccc}
 \{1,1,1\} & \{-1,-1,1\} & \{1,-1,1\} \\
 \{1,1,-1\} & \{1,1,1\} & \{1,-1,1\} \\
 \{-1,1,-1\} & \{-1,1,1\} & \{1,1,1\} \\
\end{array}
\right),
\end{align}
where $H(x)$ is the unit step function. And the function $\eta_{xy}$ can thus be defined by taking the product on the torus index $i$ as,
\begin{equation}\label{eta}
\eta_{xy}\equiv  \prod_i \sigma_{xy}^i = \left(
\begin{array}{ccc}
 1 & 1 & -1 \\
 -1 & 1 & -1 \\
 1 & -1 & 1 \\
\end{array}
\right).
\end{equation}
Using above defined $\sigma_{xy}^{i}$ and $\eta_{xy}^{i}$, we can readily write the four cases of functions $\Psi(\theta_{xy})$ defined in \eqref{eqn:Psi1} and \eqref{eqn:Psi2} into a single expression as,
\begin{align}\label{Psi}
\Psi(\theta^j_{xy}) &= \eta_{xy}\left( \frac{1}{2} \psi^{(0)}(\sigma_{xy}^{i}\theta^j_{xy})+\frac{1}{2}\psi^{(0)}(1-\sigma_{xy}^{i}\theta^j_{xy})+\gamma_E-\log(t^j+\bar t^j)\right) ,
\end{align}
where $\psi^{(0)}(z)$ is called the digamma function defined as the derivative of the logarithm of the gamma function. The successive derivatives of the $\log\Gamma (z)$ yield the polygamma function $\psi^{(n)}(z)$ as,
\begin{align}\label{polygamma}
\psi^{(n-1)}(z) = \frac{d^{(n)}}{dz^{(n)}}\log\Gamma(z) ,
\end{align}
with the following properties,
\begin{align}\label{eq:property}
\frac{d }{dz}\psi^{(0)}(\pm z) &= \pm \psi^{(1)}(\pm z), \nonumber\\
\frac{d }{dz}\psi^{(0)}(1 \pm z) &= \pm \psi^{(1)}(1 \pm z).
\end{align}
Similarly, the derivative $\Psi'(\theta^j_{xy}) = \frac{d\Psi(\theta^j_{xy})}{d \theta^j_{xy}}$ can be expressed succinctly as,
\begin{align}\label{DPsi}
\Psi'(\theta^j_{xy}) & = \eta_{xy}\sigma_{xy}^{i} \left( \frac{1}{2} \psi^{(1)}(\sigma_{xy}^{i}\theta^j_{xy})+\frac{1}{2}\psi^{(1)}(1-\sigma_{xy}^{i}\theta^j_{xy})\right),
\end{align}
where we have utilized the property \eqref{eq:property} and have neglected the contribution of the $t$-moduli.

Lastly, by making use of appropriate Kronecker deltas and defining $u^4\equiv s$, we can express the various cases of the first and second derivatives of the angles as,
\begin{align}\label{derivative-angles}
{\theta}^{i,m}_{xy} \equiv (u^m+\bar u^m)\,\frac{\partial \theta^i_{xy}}{\partial u^m} = (-1)^{\delta _{m,4}} (-1)^{\delta _{i,j}}\frac{\sin (2 \pi  \theta^i)}{4\pi } \bigg|^x_y , \nonumber \\
 i={1,2,3}; \quad m={1,2,3,4.}
\end{align}
\begin{align}
{\theta}^{i,mn}_{xy} &\equiv (u^m+\bar u^m)(u^n+\bar u^n)\,\frac{\partial^2 \theta^i_{xy}}{\partial u^m\partial\bar u^n} \nonumber \\
&= \delta _{m,n}\frac{\sin (4 \pi  \theta^i) +(-1)^{(1-\delta _{4,m}) (1-\delta _{i,m})}4 \sin (2 \pi  \theta^i)}{16\pi } \bigg|^x_y \nonumber \\
&\quad + (1-\delta _{m,n}) (-1)^{(1-\delta _{4,m}) (1-\delta _{4,n}) (\delta _{i,m}+\delta _{i,n})} (-1)^{1-\delta _{i,m}-\delta _{i,n}} \frac{\sin (4 \pi  \theta^i)}{16\pi } \bigg|^x_y \nonumber \\
& \qquad \qquad \qquad \qquad \qquad\qquad\qquad\qquad\quad i={1,2,3}; \quad m,n={1,2,3,4}.
\end{align}

Utilizing above results while ignoring the CP-violating phases $\gamma^i$, the gaugino masses, trilinear coupling \eqref{tri-coupling} and sleptons and squarks mass-squared \eqref{slepton-mass} parameters are obtained as follows,
\begin{align}\label{Gauginos123}
M_1 & = m_{3/2} (1.02057 \Theta _1-0.0583182 \Theta _2+0.653164 \Theta _3-0.116636 \Theta _4),\nonumber\\
M_2 & = m_{3/2} (0.96225 \Theta _2-0.7698 \Theta _4) ,\nonumber\\
M_3 & = m_{3/2} (1.12263 \Theta _1-0.160375 \Theta _2+0.44905 \Theta _3-0.32075 \Theta _4) ,\nonumber\\
A_0 & \equiv A_{abc}=A_{acb}= m_{3/2}\Big(-1.65515 \Theta_1-0.0769026 \Theta_2+0.210438 \Theta_3-0.210438 \Theta_4 \Big) ,\nonumber\\
m^2_{L}& \equiv m^2_{ab} = m_{3/2}{}^2 \Big(1-0.691918 \Theta_1{}^2-0.251923 \Theta_1 \Theta_2+1.38697 \Theta_1 \Theta_3+0.297487 \Theta_1 \Theta_4\nonumber\\
&\quad\quad\quad -0.730464 \Theta_2{}^2+1.38697 \Theta_2 \Theta_3+1.05521 \Theta_2 \Theta_4-2.46331 \Theta_3{}^2\nonumber\\
&\quad\quad\quad -0.104624 \Theta_3 \Theta_4-1.61046 \Theta_4{}^2\Big) ,\nonumber\\
m^2_{R}& \equiv m^2_{ac}  =m_{3/2}{}^2 \Big(1+0.672202 \Theta_1{}^2+0.024956 \Theta_1 \Theta_2-2.83616 \Theta_1 \Theta_3+0.335005 \Theta_1 \Theta_4\nonumber\\
&\quad\quad\quad -2.02285 \Theta_2{}^2-0.113933 \Theta_2 \Theta_3-1.68771 \Theta_2 \Theta_4+0.712303 \Theta_3{}^2\nonumber\\
&\quad\quad\quad -0.122343 \Theta_3 \Theta_4+0.588437 \Theta_4{}^2\Big).
\end{align}
All above results are subject to the constraint,
\begin{align}\label{constraint}
 \Theta_1^2+ \Theta_2^2+ \Theta_3^2+\Theta_4^2=1.
\end{align}

In Appendix \ref{Appendix}, we also compute the soft terms for the model with exact gauge coupling unification that was previously studied in ref.~\cite{Chen:2007zu}.

\section{Yukawa couplings}\label{sec:Yukawa}

Yukawa couplings arise from open string world-sheet instantons that connect three D-brane intersections \cite{Aldazabal:2000cn}. Intersecting D6-branes at angles wrap 3-cycles on the compact space $\mathbf{T^6}= \mathbf{T^2} \times \mathbf{T^2} \times \mathbf{T^2}$. For instance in the case of three stacks of D-branes wrapping on a $\mathbf{T^2}$ the 3-cycles can be represented by the wrapping numbers in a vector form as:
\begin{eqnarray} \label{eq:branes}
&& [\Pi_a]=(n_a, l_a) \rightarrow  z_a = R(n_a + \tau l_a) \cdot x_a,
\nonumber \\
&& [\Pi_b]=(n_b, l_b) \rightarrow  z_b = R(n_b + \tau l_b) \cdot x_b,
\nonumber \\
&& [\Pi_c]=(n_c, l_c) \rightarrow  z_c = R(n_c + \tau l_c) \cdot x_c,
\end{eqnarray}
where $\tau$ is the complex structure parameter, $(n_a, l_a)\in \mathbb{Z}^2$ is the wrapped 1-cycle, $x\in \mathbb{R}$ and $z_a\in \mathbb{C}$ respective to the brane $a$.
The triangles bounded by the triplet of D-branes $(z_a, z_b, z_c)$ will contribute to the Yukawa couplings \cite{Cremades:2003qj}. A  \textit{closer} condition,
\begin{equation}
z_a+z_b+z_c = 0, \label{close}
\end{equation}
ensures that triangles are actually formed by the three branes. The Diophantine equation \eqref{eq:branes} together with the closer condition can be solved to get the following solution:
\begin{eqnarray}
&&x_a = \frac{I_{bc}}{d} x, \nonumber \\
&&x_b = \frac{I_{ca}}{d} x, \quad x=x_0+l,  \nonumber \\
&&x_c = \frac{I_{ab}}{d} x,
\end{eqnarray}
where $I_{ab}$ is the intersection number, $d=g.c.d.(I_{ab}, I_{bc}, I_{ca})$ is the greatest common divisor of the intersection numbers, $l\in\mathbb{Z}$ arises from triangles connecting different points in the covering space $\mathbf{T^6}$ but the same points under the lattice $\mathbf{T^2}$ of the triangles and $x_0\in \mathbb{R}$ depends on the relative positions of the branes and the particular triplet $(i, j, k)$ of intersection points,
\begin{eqnarray}\label{eq:ijk}
i=0,1,\cdots, |I_{ab}|-1, \nonumber \\
j=0,1,\cdots, |I_{bc}|-1, \nonumber \\
k=0,1,\cdots, |I_{ca}|-1,
\end{eqnarray}
such that $x_0$ can be written as
\begin{equation}
x_0(i,j,k) = \frac{i}{I_{ab}}+ \frac{j}{I_{bc}} + \frac{k}{I_{ca}}.
\end{equation}
Relaxing the condition that all branes intersect at the origin, we can introduce brane shifts $\epsilon_{\alpha}$, $\alpha=a,b,c$ to write a general expressions for $x_0$ as
\begin{equation}
x_0(i,j,k) = \frac{i}{I_{ab}}+ \frac{j}{I_{bc}} + \frac{k}{I_{ca}} + \frac{d ( I_{ab}\epsilon_c + I_{ca}\epsilon_b + I_{ab}\epsilon_a
)}{I_{ab} I_{bc} I_{ca}} ,\label{x0shift}
\end{equation}
where we can absorb these three parameters into only one as,
\begin{equation}
\epsilon=\frac{I_{ab} \epsilon_c + I_{ca} \epsilon_b + I_{bc}
\epsilon_a}{I_{ab} I_{bc} I_{ca}}.
\end{equation}
This is obvious due to the reparametrization invariance in $\mathbf{T}^2$ since we can always choose two branes to intersect at the origin and the only remaining freedom left is the shift of third brane. The formula of the areas of the triangles can then be expressed using \eqref{x0shift} as,
\begin{eqnarray}
&& A(z_a, z_b) = \frac{1}{2} \sqrt{|z_a|^2|z_b|^2-(\mathrm{Re} z_a
\bar{z}_b)^2} \nonumber \\
\rightarrow && A_{ijk}(l) = \frac{1}{2} (2\pi)^2 A |I_{ab} I_{bc}
I_{ca}| (\frac{i}{I_{ab}} + \frac{j}{I_{bc}} + \frac{k}{I_{ca}} +
\epsilon + l)^2,
\end{eqnarray}
where $A$ is the K\"ahler structure of the torus. Finally, the Yukawa coupling for the three states localized at the intersections indexed by $(i,j,k)$ is given as,
\begin{equation}
Y_{ijk}= h_{qu} \sigma_{abc} \sum_{l \in Z}
\exp(-\frac{A_{ijk}(l)}{2 \pi \alpha'}),
\end{equation}
where the real phase $\sigma_{abc} = {\rm sign}(I_{ab} I_{bc} I_{ca})$ comes from the full instanton contribution  \cite{Cremades:2003qj} and $h_{qu}$ is due to quantum correction as discussed in \cite{Cvetic:2003ch}. For the ease of numerical computation real modular theta function is used to re-express the summation as
\begin{equation}
\vartheta \left[\begin{array}{c} \delta \\ \phi
\end{array} \right] (t) = \sum_{l\in\mathbf{Z}} e^{-\pi t (\delta+l)^2}
e^{2l\pi i(\delta + l) \phi}, \label{Rtheta}
\end{equation}
where the corresponding parameters are related as,
\begin{eqnarray}
&&\delta = \frac{i}{I_{ab}} + \frac{j}{I_{bc}} + \frac{k}{I_{ca}}
+ \epsilon, \nonumber\\
&& \phi=0, \nonumber\\
&& t=\frac{A}{\alpha'} |I_{ab} I_{bc} I_{ca}|.
\end{eqnarray}
Notice that the theta function $\vartheta$ is real, however $t$ can be complex while $\phi$ is an overall phase.

\subsection{Adding a $B$-field and Wilson lines}
Strings being one dimensional naturally couple to a 2-form $B$-field in addition to the metric. To incorporate the turning \emph{on} of this $B$-field leads to a \emph{complex} K\"ahler structure of the compact space $\mathbf{T^2}$ such that,
\begin{equation}
J=B+iA,
\end{equation}
and the otherwise real parameter $t$ is changed to a complex parameter $\kappa$ as,
\begin{equation}
\kappa = \frac{J}{\alpha'} |I_{ab} I_{bc} I_{ca}|.
\end{equation}
Secondly, we can also add Wilson lines around the compact directions wrapped by the D-branes. However, to avoid breaking any gauge symmetry Wilson lines must be chosen corresponding to group elements in the centre of the gauge group, i.e., a phase \cite{Cremades:2003qj}. For a triangle formed by three D-branes $a$, $b$ and $c$ each wrapping a different 1-cycle inside of $\mathbf{T}^2$, the Wilson lines can be given by the corresponding phases $\exp (2\pi i \theta_a)$, $\exp (2\pi i \theta_b)$, and $\exp (2\pi i \theta_c)$ respectively. The total phase picked up by an open string sweeping such triangle will depend upon the relative longitude of each segment, determined by the intersection points:
\begin{equation}
e^{2\pi i x_a \theta_a}  e^{2\pi i x_b \theta_b} e^{2\pi i x_c
\theta_c} = e^{2\pi i x(I_{bc} \theta_a + I_{ca} \theta_b + I_{ab}
\theta_c)}.
\end{equation}
In general, considering both a $B$-field as well as Wilson lines we get a complex theta function as
\begin{equation}
\vartheta \left[\begin{array}{c} \delta \\ \phi
\end{array} \right] (\kappa) = \sum_{l\in\mathbf{Z}} e^{\pi i \kappa
(\delta+l)^2 } e^{2 \pi i(\delta + l) \phi}, \label{Ctheta}
\end{equation}
where
\begin{align}
\delta &= \frac{i}{I_{ab}} + \frac{j}{I_{bc}} + \frac{k}{I_{ca}} + \epsilon, \nonumber\\
\phi &=I_{ab} \theta_c + I_{bc} \theta_a + I_{ca} \theta_b, \nonumber\\
\kappa &=\frac{J}{\alpha'} |I_{ab} I_{bc} I_{ca}|.
\end{align}

\subsection{O-planes and non-prime intersection numbers}
To cancel the RR-tadpoles we need to introduce the orientifold O-planes that are objects of negative tension. In addition for each D-brane $a$, we must include its mirror image $a'$ under $\Omega R$. Such mirror branes will in general wrap a different cycle $\Pi_{a*}$, related to $\Pi_{a}$ by the action of $R$ on the homology of the torus. Consequently we also need to include the triangles formed by either of the branes or their images. As an example the Yukawa coupling from the branes $a$, $b'$, and $c$ will depend on the parameters $I_{ab'}$, $I_{b'c}$, and $I_{ca}$, where the primed indexes are independent of the unprimed ones.

Furthermore, the three intersection numbers may not be coprime in general. Therefore, to avoid overcounting we need to involve the $g.c.d.$ of the intersection numbers as $d = g.c.d.(I_{ab}, I_{bc}, I_{ca})$.

Finally, to ensure that triangles are bounded by D-branes, the intersection indices must satisfy the following condition \cite{Cremades:2003qj}
\begin{equation}
i+j+k = 0 \ {\rm~mod} \ d.  \label{closetriA}
\end{equation}

\subsection{The general formula of Yukawa couplings}
Therefore, the most general formula for Yukawa couplings for D6-branes wrapping a compact $\mathbf{T^2} \times \mathbf{T^2} \times
\mathbf{T^2}$ space can be written as,
compact space  as
\begin{equation}
Y_{\{ijk\}}=h_{qu} \sigma_{abc} \prod_{r=1}^3 \vartheta
\left[\begin{array}{c} \delta^{(r)}\\ \phi^{(r)}
\end{array} \right] (\kappa^{(r)}),
\end{equation}
where
\begin{equation}
\vartheta \left[\begin{array}{c} \delta^{(r)}\\ \phi^{(r)}
\end{array} \right] (\kappa^{(r)})=\sum_{l_r\in\mathbf{Z}} e^{\pi
i(\delta^{(r)}+l_r)^2 \kappa^{(r)}} e^{2\pi i(\delta^{(r)}+l_r)
\phi^{(r)}},   \label{Dtheta}
\end{equation}
with $r=1,2,3$ denoting the three 2-tori. And the input parameters are defined by
\begin{eqnarray}
\nonumber
&&\delta^{(r)} = \frac{i^{(r)}}{I_{ab}^{(r)}} +
\frac{j^{(r)}}{I_{ca}^{(r)}} + \frac{k^{(r)}}{I_{bc}^{(r)}} +
\frac{d^{(r)} ( I_{ab}^{(r)} \epsilon_c^{(r)} + I_{ca}^{(r)}
\epsilon_b^{(r)} + I_{bc}^{(r)} \epsilon_a^{(r)}
)}{I_{ab} I_{bc} I_{ca}} + \frac{s^{(r)}}{d^{(r)}}, \\ \nonumber
&&\phi^{(r)} = \frac{I_{bc}^{(r)} \theta_a^{(r)} + I_{ca}^{(r)}
\theta_b^{(r)} + I_{ab}^{(r)} \theta_c^{(r)}}{d^{(r)}}, \\
&&\kappa^{(r)} = \frac{J^{(r)}}{\alpha'} \frac{|I_{ab}^{(r)}
I_{bc}^{(r)} I_{ca}^{(r)}|}{(d^{(r)})^2}.
\label{eqn:Yinput}
\end{eqnarray}

The theta function defined in \eqref{Ctheta} is in general complicated to evaluate numerically. However, for the special case without $B$-field, defining $J'= -iJ = A$ and $\kappa'= -i\kappa$ the $\vartheta$ function takes a more manageable form,
\begin{eqnarray}
&&\vartheta \left[\begin{array}{c} \delta \\ \phi
\end{array} \right] (\kappa') = \sum_{l\in\mathbf{Z}} e^{-\pi
\kappa' (\delta+l)^2 } e^{2 \pi i(\delta + l) \phi}, \nonumber\\
\stackrel{\textrm{redefine}}{\longrightarrow} &&\vartheta
\left[\begin{array}{c} \delta \\ \phi \end{array} \right] (\kappa)
= e^{-\pi \kappa \delta^2} e^{2\pi i \delta \phi} \vartheta_3 (\pi
(\phi+ i \kappa \delta), e^{-\pi \kappa}), \label{Ntheta}
\end{eqnarray}
in terms of $\vartheta_3$, the Jacobi theta function of the third kind.

\section{Semi-realistic Yukawa textures}\label{sec:Masses}
\begin{figure}[thb]
\centering\centering
\includegraphics[width=\textwidth]{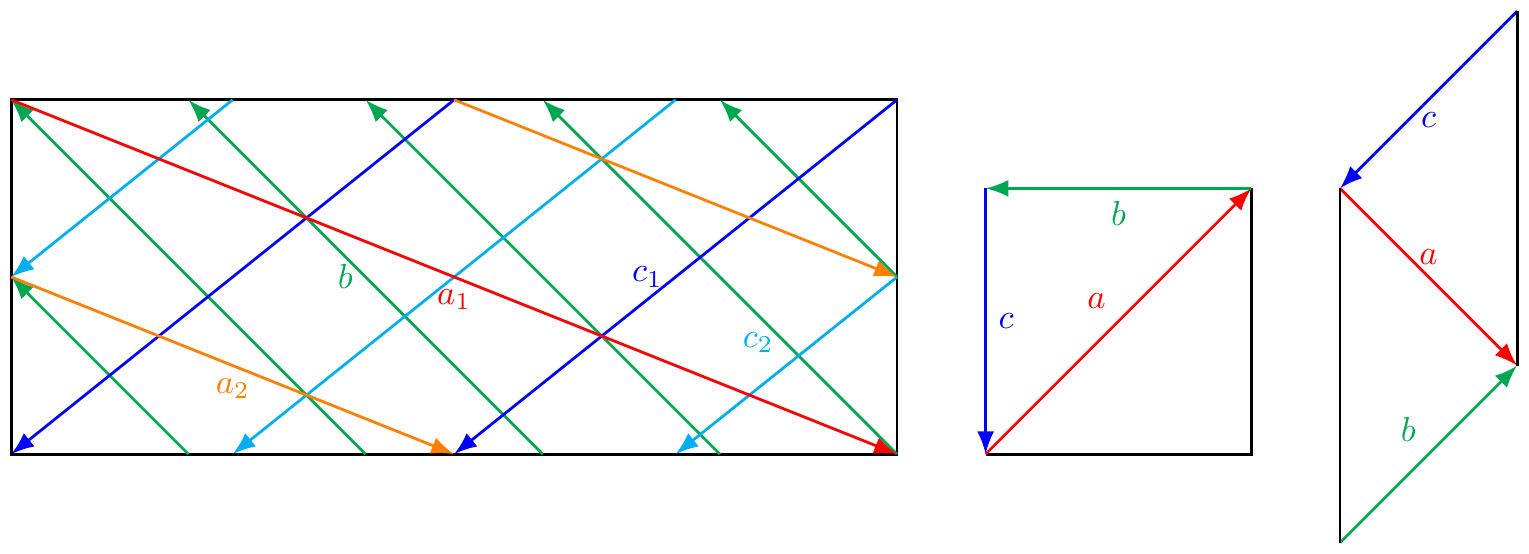}
\caption{Brane configuration for the three 2-tori where the third 2-torus is tilted. Fermion mass hierarchies result from the intersections on the first 2-torus.}  \label{3tori}
\end{figure}

Yukawa matrices for the Model 16 depicted in table \ref{tab:16} are of rank 3 and the three intersections required to form the disk diagrams for the Yukawa couplings all occur on the first torus as shown in figure~\ref{3tori}. The second and third tori only contribute an overall constant that has no effect in computing the fermion mass ratios. Thus, it is sufficient for our purpose to only focus on the first torus to nearly reproduce the correct the masses of the standard model fermions.

\subsection{Mass matrices from 3-point functions}
For the wrapping numbers listed in table~\ref{tab:16}, the intersection numbers on each torus are given as,
\begin{equation}\label{eq:Intersections}
\begin{tabular}{l l l l l}
$I_{ab}^{(1)}=3$, & $I_{ab}^{(2)}=1$, & $I_{ab}^{(3)}=1$,\\
$I_{bc}^{(1)}=9$, & $I_{bc}^{(2)}=1$, & $I_{bc}^{(3)}=0$, \\
$I_{ca}^{(1)}=3$, & $I_{ca}^{(2)}=1$, & $I_{ca}^{(3)}=1$. \\
\end{tabular}
\end{equation}
As the intersection numbers are not coprime, we define the greatest common divisor,
$d^{(1)}=g.c.d.(I_{ab}^{(1)},I_{bc}^{(1)},I_{ca}^{(1)})=3$. Thus, the arguments of the modular theta function as defined in \eqref{eqn:Yinput} can be written as,
\begin{eqnarray}
\delta^{(1)} &=& \frac{i^{(1)}}{3} + \frac{j^{(1)}}{9} + \frac{k^{(1)}}{3} + \frac{\epsilon_c^{(1)} + 3\epsilon_a^{(1)} + \epsilon_b^{(1)}}{9} + \frac{s^{(1)}}{3},  \label{eq:delta1} \\
\phi^{(1)} &=& 3 \theta_a^{(1)} + \theta_b^{(1)} + \theta_c^{(1)},  \\
\kappa^{(1)} &=& \frac{9J^{(1)}}{\alpha'}, \label{eq:kappa}
\end{eqnarray}
and recalling \eqref{eq:ijk}, we have $i=\{0,\dots ,2\}$, and $k=\{0,\dots ,2\}$ and $j=\{0,\dots ,8\}$ which respectively index the left-handed fermions, the right-handed fermions and the Higgs fields. Clearly, there arise nine Higgs fields from the $bc$ sector.

The second-last term in the right side of \eqref{eq:delta1} can be used to redefine the shift on each torus as
\begin{eqnarray}
&&\epsilon^{(1)} \equiv \frac{\epsilon_c^{(1)} + 3\epsilon_a^{(1)} + \epsilon_b^{(1)}}{9}. \label{shifts}
\end{eqnarray}
It can be noted from \eqref{eq:Intersections} that the intersection numbers on the second and third tori are either one or zero whose effect on the Yukawa couplings will be an over-all constant. The selection rule for the occurrence of a trilinear Yukawa for a given set of indices is given as,
\begin{equation}\label{selection-rule}
i^{(1)} + j^{(1)} + k^{(1)} = 0\; \mathrm{ mod }\; 3.
\end{equation}
Then, the mass matrices will take the following form for the specific values of $s^{(1)}$:

\begin{equation}
\mathcal{M}_{s^{(1)}=0}   \sim  \begin{pmatrix}
 v_{1} \a+v_{4} \d+v_{7} \g & v_{9} \c+v_{3} \f+v_{6} \q & v_{5} \b+v_{8} \e+v_{2} \p \\
 v_{9} \c+v_{3} \f+v_{6} \q & v_{5} \b+v_{8} \e+v_{2} \p & v_{1} \a+v_{4} \d+v_{7} \g \\
 v_{5} \b+v_{8} \e+v_{2} \p & v_{1} \a+v_{4} \d+v_{7} \g & v_{9} \c+v_{3} \f+v_{6} \q
\end{pmatrix},\label{eq:s0}
\end{equation}

\begin{equation}
\mathcal{M}_{s^{(1)}=-i}   \sim \left(
\begin{array}{ccc}
 v_{1} \a+v_{4} \d+v_{7} \g & v_{3} \c+v_{6} \f+v_{9} \q & v_{2} \b+v_{5} \e+v_{8} \p \\
 v_{9} \c+v_{3} \f+v_{6} \q & v_{8} \b+v_{2} \e+v_{5} \p & v_{7} \a+v_{1} \d+v_{4} \g \\
 v_{5} \b+v_{8} \e+v_{2} \p & v_{4} \a+v_{7} \d+v_{1} \g & v_{6} \c+v_{9} \f+v_{3} \q \\
\end{array}
\right),\label{eq:s-i}
\end{equation}

\begin{equation}
\mathcal{M}_{s^{(1)}=-k}   \sim  \left(
\begin{array}{ccc}
 v_{1} \a+v_{4} \d+v_{7} \g & v_{9} \c+v_{3} \f+v_{6} \q & v_{5} \b+v_{8} \e+v_{2} \p \\
 v_{3} \c+v_{6} \f+v_{9} \q & v_{8} \b+v_{2} \e+v_{5} \p & v_{4} \a+v_{7} \d+v_{1} \g \\
 v_{2} \b+v_{5} \e+v_{8} \p & v_{7} \a+v_{1} \d+v_{4} \g & v_{6} \c+v_{9} \f+v_{3} \q \\
\end{array}
\right),\label{eq:s-k}
\end{equation}

\begin{equation}
\mathcal{M}_{s^{(1)}=-j/3}   \sim  \left(
\begin{array}{ccc}
 v_{1} \a+v_{4} \a+v_{7} \a & v_{3} \d+v_{6} \d+v_{9} \d & v_{2} \g+v_{5} \g+v_{8} \g \\
 v_{3} \d+v_{6} \d+v_{9} \d & v_{2} \g+v_{5} \g+v_{8} \g & v_{1} \a+v_{4} \a+v_{7} \a \\
 v_{2} \g+v_{5} \g+v_{8} \g & v_{1} \a+v_{4} \a+v_{7} \a & v_{3} \d+v_{6} \d+v_{9} \d \\
\end{array}
\right),
\label{eq:s-j3}
\end{equation}	
where $v_i = \left\langle H_{i} \right\rangle$ and
the Yukawa couplings
$A$, $B$, $C$, $D$, $E$, $F$, $G$, $P$, and $Q$ are given by
\begin{align}\label{eq:3couplings}
A \equiv \vartheta \left[\begin{array}{c}
\epsilon^{(1)}\\ \phi^{(1)} \end{array} \right]
(\frac{9J^{(1)}}{\alpha'}), \quad\quad
B &\equiv  \vartheta \left[\begin{array}{c}
\epsilon^{(1)}+\frac{1}{9}\\  \phi^{(1)} \end{array} \right]
(\frac{9J^{(1)}}{\alpha'}), \quad
C \equiv \vartheta \left[\begin{array}{c}
\epsilon^{(1)}+\frac{2}{9}\\ \phi^{(1)} \end{array} \right]
(\frac{9J^{(1)}}{\alpha'}),  \nonumber\\
D \equiv  \vartheta \left[\begin{array}{c}
\epsilon^{(1)}+\frac{1}{3}\\  \phi^{(1)} \end{array} \right]
(\frac{9J^{(1)}}{\alpha'}),  \quad
E &\equiv \vartheta \left[\begin{array}{c}
\epsilon^{(1)}+\frac{4}{9}\\ \phi^{(1)} \end{array} \right]
(\frac{9J^{(1)}}{\alpha'}),  \quad
F \equiv  \vartheta \left[\begin{array}{c}
\epsilon^{(1)}-\frac{4}{9}\\  \phi^{(1)} \end{array} \right]
(\frac{9J^{(1)}}{\alpha'}), \nonumber\\
G \equiv \vartheta \left[\begin{array}{c}
\epsilon^{(1)}-\frac{1}{3}\\ \phi^{(1)} \end{array} \right]
(\frac{9J^{(1)}}{\alpha'}), \quad
P &\equiv \vartheta \left[\begin{array}{c}
\epsilon^{(1)}-\frac{2}{9}\\ \phi^{(1)} \end{array} \right]
(\frac{9J^{(1)}}{\alpha'}), \quad
Q \equiv  \vartheta \left[\begin{array}{c}
\epsilon^{(1)}-\frac{1}{9}\\  \phi^{(1)} \end{array} \right]
(\frac{9J^{(1)}}{\alpha'}).
\end{align}

The cases \eqref{eq:s0}, \eqref{eq:s-i} and \eqref{eq:s-k} have a similar structure whereas the last case \eqref{eq:s-j3} appears to forbid three different real eigenvalues. Thus, we only choose the case $s^{(1)}=-i$ as a representative scenario for the first three cases and will ignore the last possibility. The mass matrices for up quarks, down quarks and charged leptons are respectively given as follows:

\begin{equation}
(M_{u})_{ij}   \sim  \left(
\begin{array}{ccc}
 H_{u}^{1} \a_{u} + H_{u}^{4} \d_{u} + H_{u}^{7} \g_{u} & H_{u}^{3} \c_{u} + H_{u}^{6} \f_{u} + H_{u}^{9} \q_{u} & H_{u}^{2} \b_{u} + H_{u}^{5} \e_{u} + H_{u}^{8} \p_{u} \\
 H_{u}^{9} \c_{u} + H_{u}^{3} \f_{u} + H_{u}^{6} \q_{u} & H_{u}^{8} \b_{u} + H_{u}^{2} \e_{u} + H_{u}^{5} \p_{u} & H_{u}^{7} \a_{u} + H_{u}^{1} \d_{u} + H_{u}^{4} \g_{u} \\
 H_{u}^{5} \b_{u} + H_{u}^{8} \e_{u} + H_{u}^{2} \p_{u} & H_{u}^{4} \a_{u} + H_{u}^{7} \d_{u} + H_{u}^{1} \g_{u} & H_{u}^{6} \c_{u} + H_{u}^{9} \f_{u} + H_{u}^{3} \q_{u} \\
\end{array}
\right),
\label{eq:Mu}
\end{equation}

\begin{equation}
(M_{d})_{ij}   \sim  \left(
\begin{array}{ccc}
 H_{d}^{1} \a_{d} + H_{d}^{4} \d_{d} + H_{d}^{7} \g_{d} & H_{d}^{3} \c_{d} + H_{d}^{6} \f_{d} + H_{d}^{9} \q_{d} & H_{d}^{2} \b_{d} + H_{d}^{5} \e_{d} + H_{d}^{8} \p_{d} \\
 H_{d}^{9} \c_{d} + H_{d}^{3} \f_{d} + H_{d}^{6} \q_{d} & H_{d}^{8} \b_{d} + H_{d}^{2} \e_{d} + H_{d}^{5} \p_{d} & H_{d}^{7} \a_{d} + H_{d}^{1} \d_{d} + H_{d}^{4} \g_{d} \\
 H_{d}^{5} \b_{d} + H_{d}^{8} \e_{d} + H_{d}^{2} \p_{d} & H_{d}^{4} \a_{d} + H_{d}^{7} \d_{d} + H_{d}^{1} \g_{d} & H_{d}^{6} \c_{d} + H_{d}^{9} \f_{d} + H_{d}^{3} \q_{d} \\
\end{array}
\right),
\label{eq:Md}
\end{equation}

\begin{equation}
(M_{e})_{ij}   \sim  \left(
\begin{array}{ccc}
 H_{d}^{1} \a_{e} + H_{d}^{4} \d_{e} + H_{d}^{7} \g_{e} & H_{d}^{3} \c_{e} + H_{d}^{6} \f_{e} + H_{d}^{9} \q_{e} & H_{d}^{2} \b_{e} + H_{d}^{5} \e_{e} + H_{d}^{8} \p_{e} \\
 H_{d}^{9} \c_{e} + H_{d}^{3} \f_{e} + H_{d}^{6} \q_{e} & H_{d}^{8} \b_{e} + H_{d}^{2} \e_{e} + H_{d}^{5} \p_{e} & H_{d}^{7} \a_{e} + H_{d}^{1} \d_{e} + H_{d}^{4} \g_{e} \\
 H_{d}^{5} \b_{e} + H_{d}^{8} \e_{e} + H_{d}^{2} \p_{e} & H_{d}^{4} \a_{e} + H_{d}^{7} \d_{e} + H_{d}^{1} \g_{e} & H_{d}^{6} \c_{e} + H_{d}^{9} \f_{e} + H_{d}^{3} \q_{e} \\
\end{array}
\right),
\label{eq:Me}
\end{equation}	
Notice that the two light Higgs mass eigenstates will arise from the linear combination of the VEVs of the nine Higgs fields present in the model as,
\begin{equation}
H_{u,d}= \sum \frac{v^i_{u,d}}{\sqrt{\sum (v^i_{u,d})^2}} H^i_{u,d} \,,\label{Higgseig}
\end{equation}
with $v^i_{u,d} = \langle H^i_{u,d}\rangle$.

Pati-Salam gauge symmetry is broken down to the standard model by the process of brane-splitting as schematically shown in figure~\ref{brnsplit}, where the standard model particles are localized at their respective brane intersections. The mass hierarchies of the standard model are then easily explained by the relative shifting of the brane stacks. For instance, the left-handed quarks are localized at the intersections between the stacks $\{\textcolor{red}{a_1},~\textcolor{Green}{b}\}$ while the right-handed up-type and down-type quarks are respectively localized between stacks $\{\textcolor{red}{a_1},~\textcolor{blue}{c_1}\}$ and $\{\textcolor{red}{a_1},~\textcolor{cyan}{c_2}\}$. Thus, if we shift stack $\textcolor{cyan}{c_2}$ in the orientifold by an amount $\epsilon_{\textcolor{cyan}{c2}}$ while the stack $\textcolor{blue}{c_1}$ is unshifted ($\epsilon_{\textcolor{blue}{c_1}} = 0$), then the down-type quark masses are naturally suppressed relative to the up-type quarks. Similarly, because the left-handed and the right-handed charged leptons are respectively localized at the intersection between stacks $\{\textcolor{orange}{a_2},~\textcolor{Green}{b}\}$ and stacks $\{\textcolor{orange}{a_2},~\textcolor{cyan}{c_2}\}$, the shifting of stack $\textcolor{orange}{a_2}$ by some amount $\epsilon_{\textcolor{orange}{a_2}}$ will result in the suppression of the charged lepton masses relative to the down-type quarks. Hence, the following observed mass hierarchy is a consequence of pure geometry of the internal space,
\begin{equation}
m_{u} > m_{d} > m_{e}.
\end{equation}

By running the RGE's up to unification scale, we can determine the desired mass matrices for quarks and leptons. For example, considering $\tan\beta\equiv v_u/v_d = 50$,
the CKM matrix at the unification scale $\mu=M_X$ has been determined as \cite{Fusaoka:1998vc,Ross:2007az},
\begin{equation}
V_{\rm CKM} = \left(\begin{array}{ccc}
0.9754 & 0.2205 & -0.0026i  \\
-0.2203e^{0.003^{\circ}i} & 0.9749 & 0.0318 \\
0.0075e^{-19^{\circ}i} & -0.0311e^{1.0^{\circ}i} & 0.9995
\end{array} \right).
\end{equation}
The diagonal mass matrices for up-type and down-type quarks are respectively denoted as $D_u$ and $D_d$, and are given as,
\begin{equation}
D_u = m_t \left(\begin{array}{ccc}
0.0000139 & 0 & 0  \\
0 & 0.00404 & 0 \\
0 & 0 & 1.
\end{array} \right),\label{QUmass}
\end{equation}
\begin{equation}
D_d = m_b \left(\begin{array}{ccc}
0.00141 & 0 & 0  \\
0 & 0.0280 & 0 \\
0 & 0 & 1.
\end{array} \right) \label{Dmass},
\end{equation}
and obey the following relations,
\begin{gather}
V_{\rm CKM} =U^u_L {U^d_L}^{\dag}, \nonumber\\
D_u =U^u_L M_u {U^u_R}^{\dag},\quad \quad D_d =U^d_L M_d {U^d_R}^{\dag},
\end{gather}
where $U^i$ are the unitary matrices and $M_u M_u^{\dag}$ and $M_d M_d^{\dag}$ are the squared mass matrices of the up and down-type quarks. Similarly, the charged leptons mass eigenvalues at $\tan\beta = 50$ are given as,
\begin{equation}
D_e = m_{\tau} \left(\begin{array}{ccc}
0.000217 & 0 & 0  \\
0 & 0.0458 & 0 \\
0 & 0 & 1.
\end{array} \right).\label{eq:mass-leptons}
\end{equation}
where we have taken the ratio $m_\tau/m_b=1.58$ from the previous study of soft terms \cite{Chen:2007zu}.

\subsection{Fitting the quark masses and mixings}

In the standard model, the quark matrices $M_u$ and $M_d$ can always be made Hermitian by suitable transformation of the right-handed fields. We consider the case that $M_d$ is very close to the diagonal matrix for down-type quark, which effectively means that $U^d_L$ and $U^d_R$ are very close to the unit matrix with very small off-diagonal terms, then
\begin{equation}
V_{\rm CKM}\simeq U^u U^{d\dag}\simeq U^u,
\end{equation}
where we have transformed away the right-handed effects and made them the same as the left-handed ones.  Thus, the mass matrix of the up-type quarks becomes,
\begin{equation}
M_u \sim V_{\rm CKM}^{\dag} D_u V_{\rm CKM}.
\end{equation}
And the absolute value of $M_u$ is given as,
\begin{equation}
|M_u| = m_t \left(
\begin{array}{ccc}
0.000265544 & 0.0010868 & 0.00746948 \\
 0.0010868 & 0.00480762 & 0.0309592 \\
 0.00746948 & 0.0309592 & 0.999004 \\
\end{array}
\right).
\label{Umass}
\end{equation}
Henceforth, we need to fit \eqref{Umass} and \eqref{Dmass} to explain the mixing and the eigenvalues of the up-type and down-type quarks by fine-tuning the coupling parameters and the Higgs VEVs in \eqref{eq:Mu} and \eqref{eq:Md}. It looks at the first glance that the solution can be easily found, but we should keep in mind that the nine parameters from the theta function controlled by the D-brane shifts and Wilson-line phases are \emph{not} independent. Examining \eqref{Umass} and \eqref{eq:s-i} it is clear that we are tightly constrained by the off-diagonal terms. For instance, consider the ratio of the terms (12) and (33),
\begin{align}
\frac{v_{3}\c+v_{6}\f+v_{9}\q}{v_{3}\q+v_{6}\c+v_{9}\f} \approx 0.001,
\end{align}
which is only possible if we have $\frac{\c}{\q}\ll 1$, $\frac{\f}{\c}\ll 1$, and $\frac{\q}{\f}\ll 1$.

Comparing \eqref{eq:Mu} and \eqref{eq:Md} with the up-type quarks matrix $M_u$ and the diagonal down-type quarks matrix $D_d$, we obtain an exact fitting by expending the nine up-type Higgs VEVs and the nine down-type Higgs VEVs respectively.
\begin{align}\label{eq:Quarks3point}
  |M_{3u}| & = |M_u|, \nonumber\\
  |M_{3d}| & = D_d.
\end{align}
Here, we have set the K\"{a}hler modulus on the first 2-torus defined in \eqref{eq:kappa} as $\kappa^{(1)} = 45$ and evaluate the couplings functions \eqref{eq:3couplings} by setting geometric brane position parameters as $\epsilon^{(1)}_{u} = 0$ and $\epsilon^{(1)}_{d} = 2/9$ which yields in an exact fitting for the following VEVs,
\begin{equation}\label{eq:VEVs}
\begin{array}{ll}
 v_u^1=0.000265535 & v_d^1=4.57419\times 10^{-9} \\
 v_u^2=0.0426383 & v_d^2=0 \\
 v_u^3=5.72205 & v_d^3=0.114556 \\
 v_u^4=0.0309592 & v_d^4=-8.59558 \times 10^{-7} \\
 v_u^5=0.0425567 & v_d^5=0.00056 \\
 v_u^6=0.00635385 & v_d^6=-0.000609616 \\
 v_u^7=0.0309592 & v_d^7=0.000161524 \\
 v_u^8=0.0273106 & v_d^8=0\\
 v_u^9=-0.0242253 & v_d^9=3.24411 \times 10^{-6} .\\
\end{array}
\end{equation}

\subsection{Fitting the charged lepton masses}
Note that the down-type quark mass matrix and the lepton mass matrix both involve the same down-type Higgs VEVs. Thus, once the parameters needed to fit the down-type mass matrix are fixed, the only freedom in calculating the charged lepton mass matrix is from the geometric position $\epsilon^{(1)}_{e}$ of each brane against the set value of the parameter $\kappa^{(1)}$. We have calculated the spectrum of mass eigenvalues for charged leptons by varying $\epsilon^{(1)}_e$ from 0 to 1 for various values of $\kappa^{(1)}$.
\begin{figure}[t]\centering
\includegraphics[width=.5\textwidth]{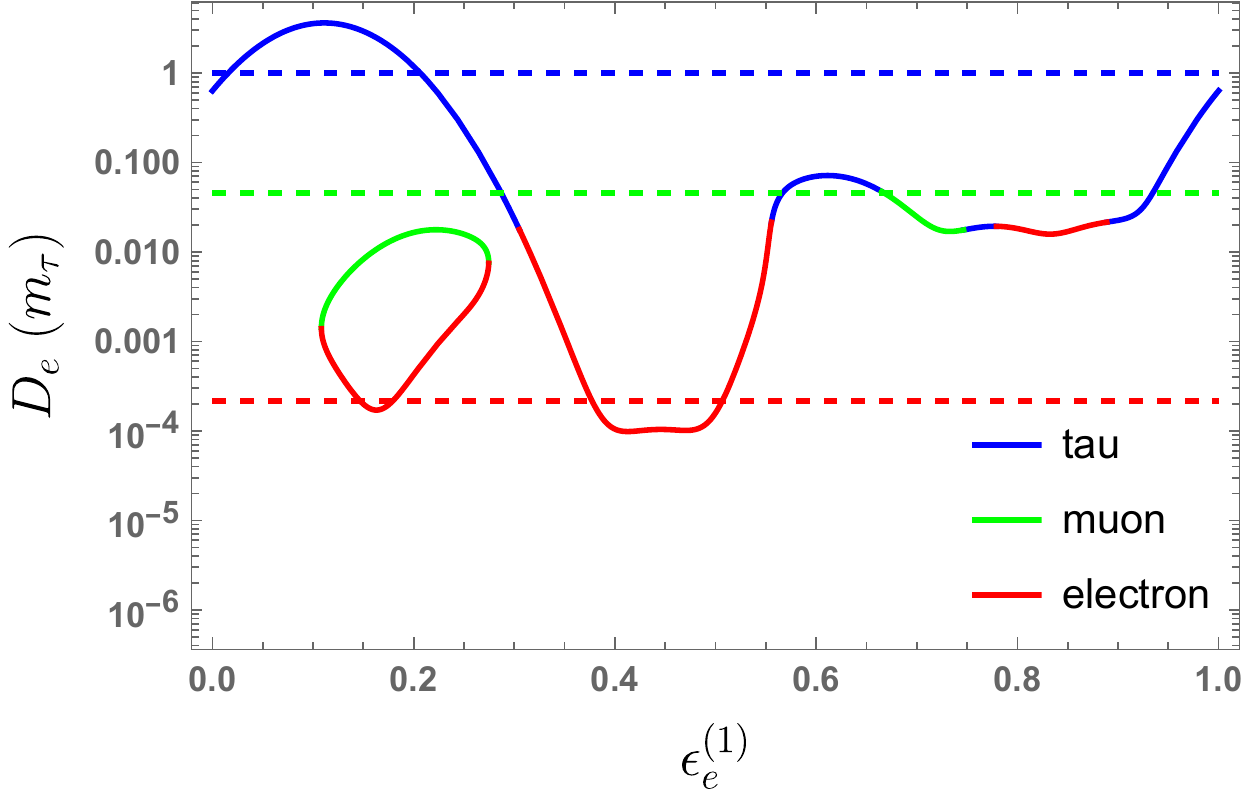}
\caption{Log plot of the spectrum of eigenvalues of the charged leptons mass as a function of brane position parameter $\epsilon^{(1)}_{e}$ for $\kappa^{(1)}=45$. The dashed colored horizontal lines correspond to the mass eigenvalues of tau, muon and electron $\{1,0.0458,0.000217\}m_\tau$ from \eqref{eq:mass-leptons}. The exact solution will only be obtained if the respective colored lines touch the dashed grid-lines for a specific value of $\epsilon^{(1)}_{e}$. The nearest-fit is obtained at $\epsilon^{(1)}_{e}=0.206558$ with eigenvalues $\{0.999993,0.0171119,0.00053237\}m_\tau$.}  \label{fig:Leptons}
\end{figure}
Figure~\ref{fig:Leptons} shows one such spectrum for the specific value of $\kappa^{(1)}=45$. It can be easily seen that the nearest match is obtained for $\epsilon^{(1)}_e=0.206558$,
\begin{equation}\label{eq:Leptons3point}
|M_{3e}|=m_\tau \left(
\begin{array}{ccc}
 0.000526961 & -0.00410568 & 8.71084\times 10^{-9} \\
 0.0000218487 & 0.0171173 & -5.03921 \times 10^{-8} \\
 -1.99023 \times 10^{-9} & 9.46941 \times 10^{-6} & 0.999993 \\
\end{array}
\right)
\end{equation}
with eigenvalues $\{0.999993,0.0171119,0.00053237\}m_\tau$, where tau lepton is fitted exactly while the muon's mass comes out to be only 37\% while the electron is about 2.45 times heavier than \eqref{eq:mass-leptons}. Notice, that these results are only at the tree-level and there could indeed be other corrections, such as those coming from higher-dimensional operators, which may contribute most greatly to the electron and muon masses since they are lighter.

\section{Yukawa couplings from 4-point functions}\label{sec:4point}

We now turn our attention to the discussion of four-point functions that affect more greatly to the masses of the lighter fermions. We are looking
for four-point interactions such as
\begin{equation} \label{4pointcases}
\phi^i_{ab} \phi^j_{ca} \phi^k_{b'c} \phi^l_{bb'} ~~\mathrm{or}~~
\phi^i_{ab} \phi^j_{ca} \phi^k_{cc'} \phi^l_{bc'}~,~\,
\end{equation}
where $\phi^i_{x y}$ are the chiral superfields at the
intersections between stack $x$ and $y$ D6-branes.
The formula for the area of a quadrilateral in terms of its angles
and two sides and the solutions of diophantine equations for
estimating the multiple areas of the quadrilaterals from non-unit
intersection numbers are given in~\cite{Abel:2003yx, Abel:2003vv}.
In addition to these formulae, there
is a more intuitive way to calculate the area for these four-sided
polygons. A quadrilateral can be always taken as the difference between two similar triangles.
Therefore, since we know the classical part is
\begin{equation}
Z_{4cl} \sim e^{-A_{quad}},
\end{equation}
it is equivalent to write \cite{Chen:2008rx},
\begin{equation}
Z_{4cl} \sim e^{-|A_{tri}-A'_{tri}|}.
\end{equation}

\begin{figure}[t]
\centering
\includegraphics[width=.5\textwidth]{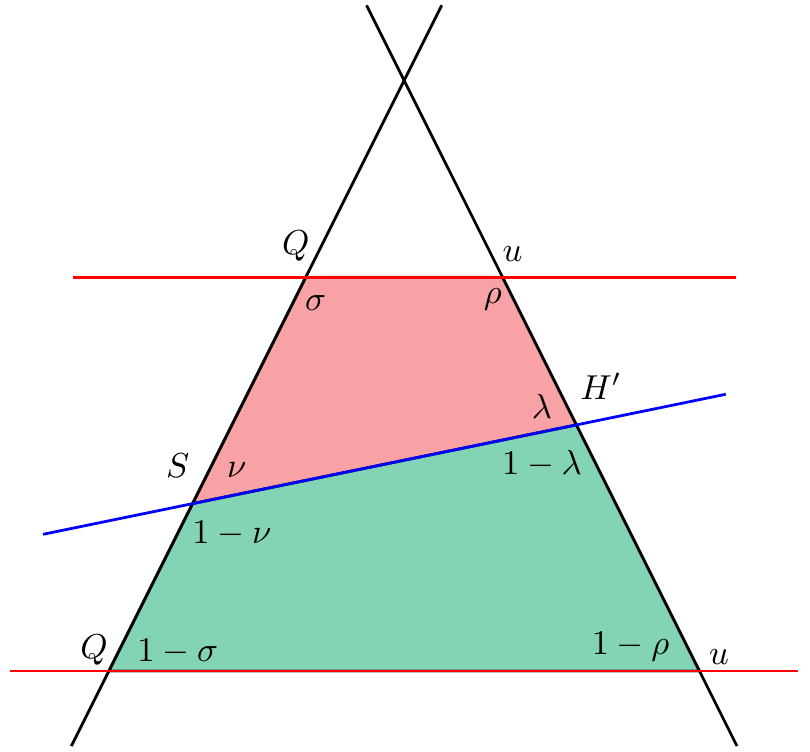}
\caption{A picture of two quadrilaterals with different field orders.
The red brane repeats in a next cycle and can still form a similar
quadrilateral with the blue brane. This coupling also contributes to
the four-point function.} \label{2-cases}
\end{figure}

Taking the absolute value of the difference reveals that there are
two cases: $A_{tri}>A'_{tri}$ and $A_{tri}<A'_{tri}$, as shown in
figure~\ref{2-cases}.  From the figure we can see the two quadrilaterals
are similar with different sizes, but the orders of the fields
corresponding to the angles are different, which is under an
interchange of $\theta \leftrightarrow 1-\theta$, $\theta=\nu,
\lambda, \rho,\sigma$. These different field orders may cause different values
for their quantum contributions. Here, we shall only consider the classical contribution from the 4-point interaction and ignore the quantum part which was shown to be further suppressed, consult \cite{Chen:2008rx} and references therein for details. Therefore, we are
able to employ the same techniques which have developed for
calculating the trilinear Yukawa couplings.

\begin{figure}[h]
\centering
\includegraphics[width=.5\textwidth]{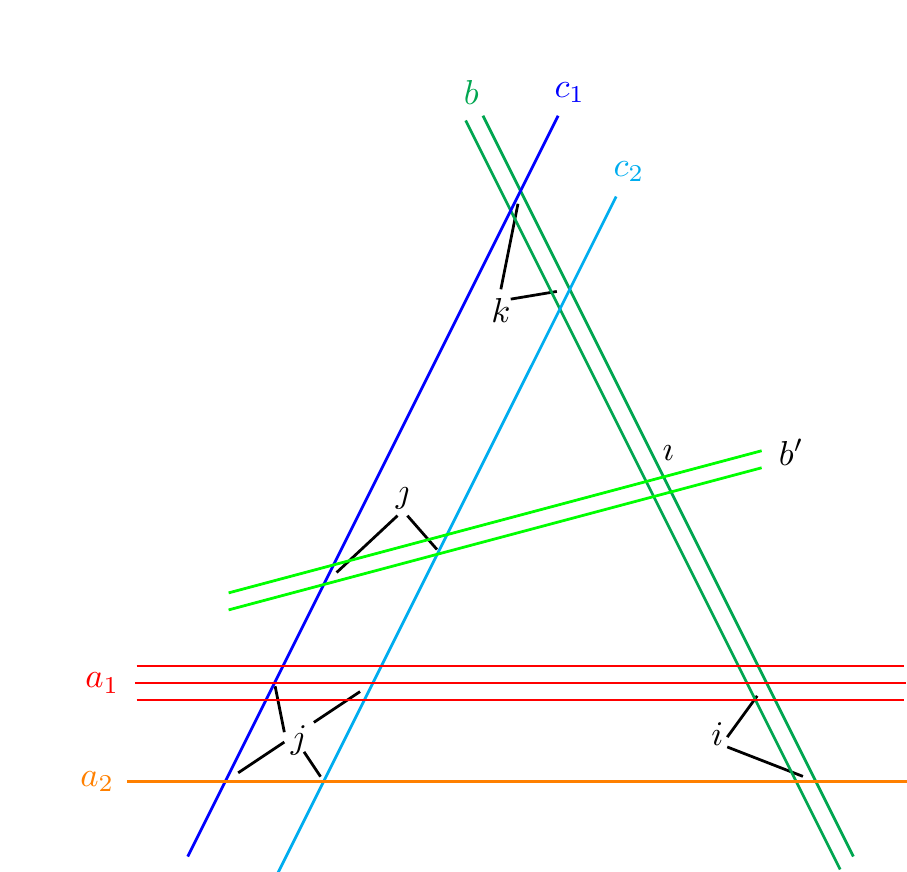}
\caption{A diagram showing the areas bounded by stacks of D-branes
which give rise to the Yukawa couplings for quarks and leptons via
world-sheet instantons.  The Yukawa couplings of the up-type
quarks are from the areas by stack $a_1$, $b$, $c_1$, the
down-type quarks by stack $a_1$, $b$, $c_2$, and the leptons by
$a_2$, $b$, $c_2$.  The four-point function corrections to the
Yukawa couplings of the up-type quarks are from the areas by stack
$a_1$, $b$, $b'$, $c_1$, the down-type quarks by stack $a_1$, $b$,
$b'$, $c_2$, and the leptons by $a_2$, $b$, $b'$, $c_2$. }
\label{brane area_bb}
\end{figure}

For a quadrilateral formed by the stacks $a$, $b$, $b'$, $c$, we can
calculate it as the difference between two triangles formed by
stacks $a$, $b$, $c$ and $b'$, $b$, $c$. In other words, they
share the same intersection $I_{bc}$.  Therefore, if we use this
method to calculate the quadrilateral area, we should keep in mind
that the intersection index $k$ for $I_{bc}$ remains the same for
a certain class of quadrilaterals when varying other intersecting
indices. Here we set indices $i$ for $I_{ab}$, $j$ for $I_{ca}$,
$\imath$ for $I_{bb'}$, and $\jmath$ for $I_{cb'}$, as shown in
figure~\ref{brane area_bb}. We may calculate the areas of the triangles
as we did in the trilinear Yukawa couplings
above~\cite{Cremades:2003qj}
\begin{eqnarray}
&&A_{ijk}=\frac{1}{2}(2\pi)^2 A_{\mathbf{T^2}}
|I_{ab}I_{bc}I_{ca}|~ \big(\frac{i}{I_{ab}} +\frac{j}{I_{ca}}
+\frac{k}{I_{bc}}
+\epsilon +l \big)^2, \nonumber \\
&&A_{\imath\jmath k}=\frac{1}{2}(2\pi)^2 A_{\mathbf{T^2}}
|I_{b'b}I_{bc}I_{cb'}|~ \big( \frac{\imath}{I_{b'b}}
+\frac{\jmath}{I_{cb'}} +\frac{k}{I_{bc}} +\varepsilon +\ell
\big)^2,  \label{Tareas}
\end{eqnarray}
where $i$, $j$, $k$ and $\imath$, $\jmath$, $k$ are using the same
selection rules as Eq. (\ref{selection-rule}). Thus, the classical
contribution of the four-point functions is given by
\begin{equation}
Z_{4cl} = \sum_{l,\ell} e^{-\frac{1}{2\pi}|A_{ijk}-A_{\imath\jmath
k}|}.
\end{equation}
Note that this formula will diverge when $A_{ijk}=A_{\imath\jmath
k}$, which is due to over-counting the zero area when the
corresponding parameters in Eq. (\ref{Tareas}) are the same.  In
such a case, $Z_{4cl} = 1+\sum_{l\neq\ell}
e^{-\frac{1}{2\pi}|A_{ijk}-A_{\imath\jmath k}|}$.  We will not
meet this special situation in our following discussion.

We will consider both types of possible interactions \eqref{4pointcases} coming from considering $b'$ or from considering $c'$ independently.

\subsection{Mass corrections from 4-point functions considering $b'$}

In the model of table~\ref{tab:16}, in addition to the intersection
numbers in Eq.~(\ref{eq:Intersections}), we have
\begin{align}\label{Intersection}
 I_{cb'}^{(1)}&= 1,    & I_{cb'}^{(2)} &= -1,   & I_{cb'}^{(3)} &= 1; \nonumber \\
 I_{bb'}^{(1)}&= 20,   & I_{bb'}^{(2)} &= 0,    & I_{bb'}^{(3)} &= -1; \nonumber \\
 I_{cc'}^{(1)}&= -4,   & I_{cc'}^{(2)} &= 0,    & I_{cc'}^{(3)} &= -1;
\end{align}
There are twenty SM singlet fields $S_L^i$ and one
Higgs-like state $H^{\prime}_{u, d}$. Similar to the Higgs
fields $H^{i}_{u, d}$, only three linear combinations of the twenty
$S_L^i$ can contribute to the four-point Yukawa couplings. Considering the following parameters with shifts $l=-k/3$ and $\ell=-k/9$ taken along the index $k$,
\begin{align}\label{deltas}
\delta &=\frac{i}{I_{ab}^{(1)}} +\frac{j}{I_{ca}^{(1)}} +\frac{k}{I_{bc}^{(1)}} +l , \nonumber\\
&= \frac{i}{3}+\frac{j}{3} ,\\
d &=\frac{\imath}{I_{cc'}^{(1)}} +\frac{\jmath}{I_{cb'}^{(1)}}+\frac{k}{I_{bc}^{(1)}} +\ell ,\nonumber\\
&= \frac{\imath}{20}+\jmath ,
\end{align}
the matrix elements $a_{i,j,\imath}$ on the first torus from the four-point functions can be expressed as
\begin{equation}
\left(
\begin{array}{ccc}
\sum _{i=0}^6 a_{1,1,3i+1} & \sum _{i=0}^5 a_{1,2,3i+3} & \sum _{i=0}^6 a_{1,3,3i+2} \\
\sum _{i=0}^5 a_{2,1,3i+3} & \sum _{i=0}^6 a_{2,2,3i+2} & \sum _{i=0}^6 a_{2,3,3i+1} \\
\sum _{i=0}^6 a_{3,1,3i+2} & \sum _{i=0}^6 a_{3,2,3i+1} & \sum _{i=0}^5 a_{3,3,3i+3} \\
\end{array}
\right),
\end{equation}
and the classical 4-point contribution to the mass matrix given as,
\begin{equation}\label{eq:4point}
Z_{4cl}=\left(
\begin{array}{ccc}
 u_1 \sum _{i=0}^6 w_{3 i+1} A_{3 i}   & u_2 \sum _{i=0}^5 w_{3 i+3} A_{3 i+2} & u_3 \sum _{i=0}^6 w_{3 i+2} A_{3 i+1} \\
 u_1 \sum _{i=0}^5 w_{3 i+3} A_{3 i+2} & u_2 \sum _{i=0}^6 w_{3 i+2} A_{3 i+1} & u_3 \sum _{i=0}^6 w_{3 i+1} A_{3 i} \\
 u_1 \sum _{i=0}^6 w_{3 i+2} A_{3 i+1} & u_2 \sum _{i=0}^6 w_{3 i+1} A_{3 i}   & u_3 \sum _{i=0}^5 w_{3 i+3} A_{3 i+2} \\
\end{array}
\right),
\end{equation}
where $u_i, w_{j}$ are the VEVs and the couplings are defined as,
\begin{equation}\label{4-point-couplings}
A_i \equiv \vartheta \left[\begin{array}{c}
\epsilon^{(1)}+\dfrac{i}{20}\\
\phi^{(1)}
\end{array}  \right]  (\frac{9J^{(1)}}{\alpha'}),\qquad i={0,\dots,|I_{bb'}^{(1)}|-1}.
\end{equation}

Since, we have already fitted the up-type quark matrix $|M_{3u}|$ exactly, so its 4-point correction should be zero,
\begin{equation}
|M_{4u}| = 0 ,
\end{equation}
which is true by setting all up-type VEVs $u_u^i$ and $w_u^i$ to be zero.
Therefore, we are essentially concerned with fitting charged leptons in such a way that corresponding corrections for the down-type quarks remain negligible.
The desired solution can be readily obtained by setting $\epsilon^{(1)}_{4e}=0$ and $\epsilon^{(1)}_{4d}=1/20$ with the following values of the VEVs,
\begin{equation}\label{eq:Wd}
\begin{array}{lll}
u_d^1=-1/27,      &  u_d^2=1,     & u_d^3=1/100 , \\
w_d^i=0, &  i={1,\dots,14}, & \\
w_d^{15}=0.172138 , & w_d^{16}=0.0858488 &   w_d^{17}=0.0247877, \\
w_d^{18}=-0.0000123476, & w_d^{19}=-0.0000513328, & w_d^{20}=-0.000123552 . \\
\end{array}
\end{equation}
The 4-point correction to the charged leptons masses is given by,
\begin{equation}\label{M4e}
|M_{4e}|=m_\tau \left(
\begin{array}{ccc}
 -0.000309961 & 0.000697239 & 0.000286827 \\
 -0.0000258237 & 0.0286827 & 0.0000836895 \\
 -0.00106232 & 0.00836895 & 6.97239 \times 10^{-6} \\
\end{array}
\right)
\end{equation}
which can be added to the matrix obtained from 3-point functions \eqref{eq:Leptons3point} as,
\begin{equation}\label{Me}
|M_{3e}|+|M_{4e}|=m_\tau \left(
\begin{array}{ccc}
 0.000217 & -0.00340844 & 0.000286835 \\
 -3.97501 \times 10^{-6} & 0.0458 & 0.0000836391 \\
 -0.00106232 & 0.00837842 & 1. \\
\end{array}
\right)
\end{equation}
that can be readily diagonalized as,
\begin{equation}
m_\tau \left(
\begin{array}{ccc}
 0.00217014 & 0 & 0 \\
 0 & 0.0457996 & 0 \\
 0 & 0 & 1. \\
\end{array}
\right)\sim D_e ~,
\end{equation}
which exactly reproduces the correct masses of all three charged leptons cf. \eqref{eq:mass-leptons}. Also, the corrections to down-type quarks are kept to almost zero by setting $\epsilon^{(1)}_{4d}=1/20$,
\begin{equation}\label{Md}
|M_{4d}|\sim 0.
\end{equation}
which preserves our previously obtained exact fit using 3-point functions \eqref{eq:Quarks3point}.

\subsection{Mass corrections from 4-point functions considering $c'$}

\begin{figure}[h]
\centering
\includegraphics[width=.5\textwidth]{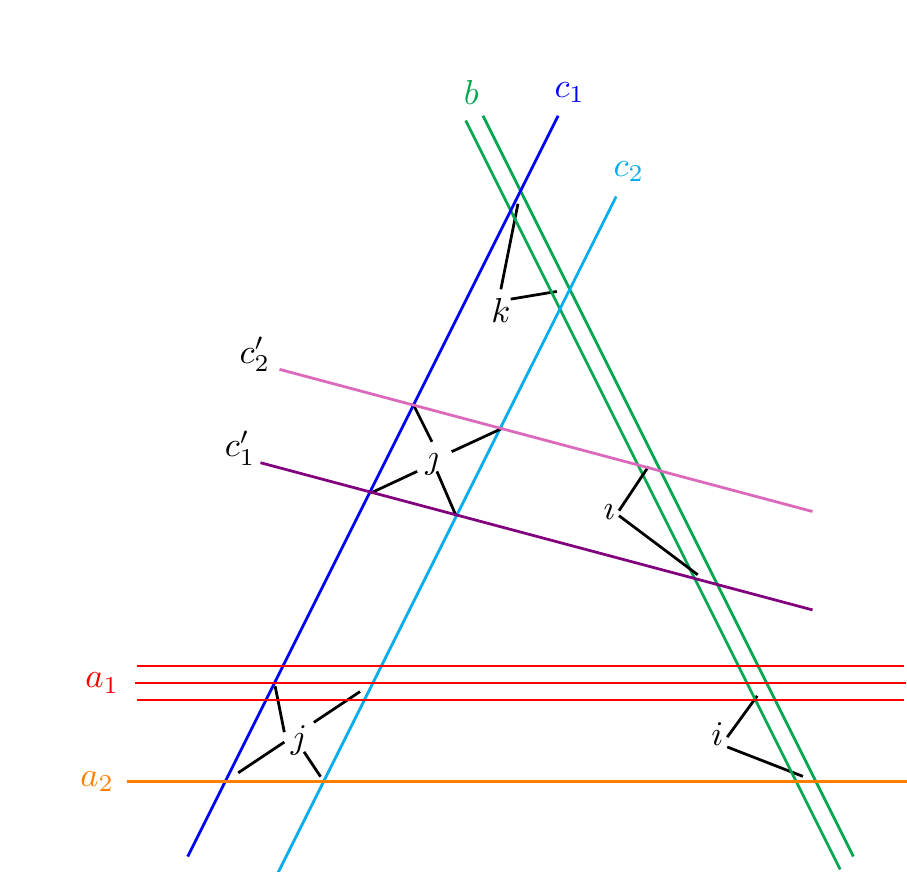}
\caption{A diagram showing the areas bounded by stacks of D-branes
which give rise to the Yukawa couplings for quarks and leptons via
world-sheet instantons.  The Yukawa couplings of the up-type
quarks are from the areas by stack $a_1$, $b$, $c_1$, the
down-type quarks by stack $a_1$, $b$, $c_2$, and the leptons by
$a_2$, $b$, $c_2$.  The four-point function corrections to the
Yukawa couplings of the up-type quarks are from the areas by stack
$a_1$, $b$, $c'$, $c_1$, the down-type quarks by stack $a_1$, $b$,
$c'$, $c_2$, and the leptons by $a_2$, $b$, $c'$, $c_2$. }
\label{brane area_cc}
\end{figure}

There are four SM singlet fields $S_L^i$ and one
Higgs-like state $H^{\prime}_{u, d}$. Similar to the Higgs
fields $H^{i}_{u, d}$, only three linear combinations of the twenty
$S_L^i$ can contribute to the four-point Yukawa couplings. Considering the following parameters with shifts $l=-k/3$ and $\ell=-k/9$ taken along the index $k$,
\begin{align}\label{deltas}
\delta &=\frac{i}{I_{ab}^{(1)}} +\frac{j}{I_{ca}^{(1)}} +\frac{k}{I_{bc}^{(1)}} +l , \nonumber\\
&= \frac{i}{3}+\frac{j}{3}, \\
d &=\frac{\imath}{I_{cc'}^{(1)}} +\frac{\jmath}{I_{cb'}^{(1)}}+\frac{k}{I_{bc}^{(1)}} +\ell ,\nonumber\\
&= -\frac{\imath}{4}+\jmath ,
\end{align}
the matrix elements $a_{i,j,\imath}$ on the first torus from the four-point functions can be expressed as,
\begin{equation}
\left(
\begin{array}{ccc}
 a_{1,1,4}+a_{1,1,1} & a_{1,2,3} & a_{1,3,2} \\
 a_{2,1,3} & a_{2,2,2} & a_{2,3,1}+a_{2,3,4} \\
 a_{3,1,2} & a_{3,2,1}+a_{3,2,4} & a_{3,3,3} \\
\end{array}
\right).
\end{equation}
and the classical 4-point contribution to the mass matrix given as,
\begin{equation}\label{eq:4point}
Z_{4cl}=\left(
\begin{array}{ccc}
 u_1 (w_4 A_1 + w_1 A_0) & u_2  w_3 A_2 & u_3 w_2 A_3 \\
 u_1 w_3 A_2 & u_2 w_2 A_3 & u_3 (w_4 A_1 + w_1 A_0) \\
 u_1 w_2 A_3 & u_2 (w_4 A_1 + w_1 A_0) & u_3 w_3 A_2 \\
\end{array}
\right),
\end{equation}
where $u_i, w_{j}$ are the VEVs and the couplings are defined as,
\begin{equation}\label{4-point-couplings}
A_i \equiv \vartheta \left[\begin{array}{c}
\epsilon^{(1)}+\dfrac{i}{4}\\
\phi^{(1)}
\end{array}  \right]  (\frac{9J^{(1)}}{\alpha'}),\qquad i={0,\dots,|I_{cc'}^{(1)}|-1}.
\end{equation}

Since, we have already fitted the up-type quark matrix $|M_{3u}|$ exactly, so its 4-point correction should be zero,
\begin{equation}
|M_{4u}| = 0,
\end{equation}
which is true by setting all up-type VEVs $u_u^i$ and $w_u^i$ to be zero.
Therefore, we are essentially concerned with fitting charged leptons in such a way that corresponding corrections for the down-type quarks remain negligible.
The desired solution can be readily obtained by setting $\epsilon^{(1)}_{4e}=1/4$ and $\epsilon^{(1)}_{4d}=3/4$ with the following values of the VEVs,
\begin{equation}\label{eq:Wd}
\begin{array}{lll}
u_d^1=-1/27,      &  u_d^2=1,     & u_d^3=1/100 , \\
w_d^{1}=1.8183 , & w_d^{2}= 0.000906372 , & \\
w_d^{3}=0.151488, & w_d^{4}=-0.000264 . & \\
\end{array}
\end{equation}
The 4-point correction to the charged leptons masses is given by,
\begin{equation}\label{M4e}
|M_{4e}|=m_\tau \left(
\begin{array}{ccc}
 -0.000309961 & 0.000697239 & 0.000286827 \\
 -0.0000258237 & 0.0286827 & 0.0000836895 \\
 -0.00106232 & 0.00836895 & 6.97239 \times 10^{-6} \\
\end{array}
\right),
\end{equation}
which can be added to the matrix obtained from 3-point functions \eqref{eq:Leptons3point} as,
\begin{equation}\label{Me}
|M_{3e}|+|M_{4e}|=m_\tau \left(
\begin{array}{ccc}
 0.000217 & -0.00340844 & 0.000286835 \\
 -3.97501\times 10^{-6} & 0.0458 & 0.0000836391 \\
 -0.00106232 & 0.00837842 & 1. \\
\end{array}
\right),
\end{equation}
that can be readily diagonalized as,
\begin{equation}
m_\tau \left(
\begin{array}{ccc}
 0.00217014 & 0 & 0 \\
 0 & 0.0457996 & 0 \\
 0 & 0 & 1. \\
\end{array}
\right)\sim D_e ~,
\end{equation}
which exactly reproduces the correct masses of all three charged leptons cf. \eqref{eq:mass-leptons}. Also, the corrections to down-type quarks are kept to almost zero by setting $\epsilon^{(1)}_{4d}=3/4$,
\begin{equation}\label{Md}
|M_{4d}|= \left(
\begin{array}{ccc}
 -8.49479\times 10^{-7} & 0.00110164 & 0 \\
 -0.0000408014 & 0 & 2.29359\times 10^{-7} \\
 0 & 0.0000229359 & 0.0000110164 \\
\end{array}
\right)\sim 0 ,
\end{equation}
which preserves our previously obtained exact fit using 3-point functions \eqref{eq:Quarks3point}.

In summary, we can correctly reproduce, the correct masses and mixings of quarks and the masses of charged leptons at the electroweak scale. Finally, by choosing suitable Majorana mass matrix for the right-handed neutrinos the suitable masses of neutrinos and their mixings can be generated by type I seesaw mechanism.

\section{Discussion and conclusion}\label{sec:conclusion}

We have studied the phenomenology of a new class of supersymmetric Pati-Salam intersecting D6-brane model on a $\mathbf{T}^6/(\mathbb{Z}_2\times \mathbb{Z}_2)$ orientifold in type IIA string theory. The defining characteristic of this new-class is that one of the wrapping numbers is 5 and models exhibit approximate gauge coupling unification. We have discussed the SM fermion masses and mixings and supersymmetry breaking soft terms in the $u$-moduli dominated case and the $u$-moduli dominant case together with the $s$-moduli turned on, where the soft terms remain independent of the Yukawa couplings and the Wilson lines. The results depend on the brane wrapping numbers as well as supersymmetry breaking parameters.

Although we are able to reproduce the extrapolated GUT-scale mass hierarchies and the mixings of the standard model fermions in our intersecting brane model on a IIA orientifold setting exactly, several issues remain to be addressed. The values of various couplings and other parameters are not determined uniquely within the model and are rather put by hand. All such parameters are the functions of open and closed string moduli. For instance the Yukawa couplings depend on the geometric position of the stacks of branes and the K\"{a}hler moduli. Once we fix these moduli, the only freedom left is the  Higgs sector, i.e. finding a specific linear combination of nine pairs of Higgs states which may be fine-tuned to yield the two Higgs eigenstates $H_u$ and $H_d$ of the MSSM.

Fixing the brane positions is thus equivalent to fixing the open string moduli. Unless these open string moduli are fixed, the low energy spectrum will contain non-chiral open string states associated to the brane positions and the Wilson lines. We do not see such scalar particles in Nature. Luckily so, otherwise they will also spoil the gauge coupling unification in the MSSM. Therefore, it is tempting to eliminate such non-chiral fields by considering intersecting D-brane models wrapping on rigid cycles. In the case of type II compactifications, $\mathbf{T}^6/(\mathbb{Z}_2\times \mathbb{Z}'_2)$ is the only known toroidal background possessing such rigid cycles, see ref.~\cite{Blumenhagen:2005tn} for details. This may be explored in a future study.

\section*{Acknowledgements}
TL is supported by the National Key Research and Development Program of China Grant No. 2020YFC2201504, by the Projects No. 11875062, No. 11947302, and No. 12047503 supported by the National Natural Science Foundation of China, as well as by the Key Research Program of the Chinese Academy of Sciences, Grant No. XDPB15. The work of AM is supported by the Zhejiang Provincial Natural Science Foundation of China (Grant No. LZ21A040001), the National Natural Science Foundation of China (Grant No. 12074344), and the Key Projects of the Natural Science Foundation of China (Grant No. 11835011). XW is supported by the National Natural Science Foundation of China (NSFC) under Grants No. 11605043.

\appendix

\section{Soft terms from susy breaking for model in ref.~\cite{Chen:2007zu}}\label{Appendix}

\begin{table}[ht]
\footnotesize
\renewcommand{\arraystretch}{1.0}
\caption{D6-brane configurations and intersection numbers for
the model on Type IIA $\mathbf{T}^6 /\mathbb{Z}_2 \times \mathbb{Z}_2$
orientifold. }
\label{MI-Numbers}
\begin{center}
\begin{tabular}{|c|c|c|c|c|c|c|c|c|c|c|c|c|}
\hline
& \multicolumn{12}{c|}{${\rm U}(4)_C\times {\rm U}(2)_L\times {\rm U}(2)_R\times {\rm USp}(2)^4$}\\
\hline   & $N$ & $(n^1,l^1)\times (n^2,l^2)\times

(n^3,l^3)$ & $n_{S}$& $n_{A}$ & $b$ & $b'$ & $c$ & $c'$& 1 & 2 & 3 & 4 \\

\hline

    $a$&  8& $(0,-1)\times (1,1)\times (1,1)$ & 0 & 0  & 3 & 0 & -3 & 0 & 1 & -1 & 0 & 0\\

    $b$&  4& $(3,1)\times (1,0)\times (1,-1)$ & 2 & -2  & - & - & 0 & 0 & 0 & 1 & 0 & -3 \\

    $c$&  4& $(3,-1)\times (0,1)\times (1,-1)$ & -2 & 2  & - & - & - & - & -1 & 0 & 3 & 0\\

\hline

    1&   2& $(1,0)\times (1,0)\times (2,0)$ & \multicolumn{10}{c|}{$\chi_1=3,~
\chi_2=1,~\chi_3=2$}\\

    2&   2& $(1,0)\times (0,-1)\times (0,2)$ & \multicolumn{10}{c|}{$\beta^g_1=-3,~
\beta^g_2=-3$}\\

    3&   2& $(0,-1)\times (1,0)\times (0,2)$& \multicolumn{10}{c|}{$\beta^g_3=-3,~
\beta^g_4=-3$}\\

    4&   2& $(0,-1)\times (0,1)\times (2,0)$ & \multicolumn{10}{c|}{}\\

\hline

\end{tabular}

\end{center}

\end{table}

Table~\ref{MI-Numbers} shows the well-studied previous model with wrapping number up to 3 with exact gauge coupling unification. This model has been extensively discussed in Refs. \cite{Chen:2007zu, Mayes:2013bda, Mayes:2019isy, Gemmill:2019kxr, Li:2014xqa}. The gaugino masses, trilinear coupling and the squared masses of sleptons and squarks for this model computed in \cite{Li:2014xqa} did not take into account the fact that the third torus is tilted. Consequently, there was some discrepancy in the trilinear coupling and the squared sleptons and squarks masses. Below we perform the computation making use of Kronecker deltas and the sign matrix $\sigma^i_{xy}$ defined in \eqref{sigmaK}, which conveniently takes into account the signs in the derivative of the angles and the $\Psi$-functions.

From table~\ref{MI-Numbers}, the complex structure moduli $U$ from \eqref{U-moduli} can be calculated as,
\begin{align}\label{eqn:U-moduli}
\{U_1,U_2,U_3\} & =  \{3 i, i, 1+i\} ,
\end{align}
and the corresponding $u$-moduli and $s$-modulus in supergravity basis from \eqref{eq:moduli} are,
\begin{align}\label{s-u-moduli}
u_i & =\{\frac{e^{-\phi_4}}{\sqrt{6} \pi },\frac{\sqrt{\frac{3}{2}} e^{-\phi_4}}{\pi },\frac{\sqrt{\frac{3}{2}} e^{-\phi_4}}{2 \pi }\}, \nonumber\\
s & =  \frac{e^{-\phi_4}}{2 \sqrt{6} \pi } .
\end{align}
Using \eqref{kingauagefun} and the values from the table~\ref{MI-Numbers}, the gauge kinetic function becomes,
\begin{align}\label{fx}
f_x & = \frac{\sqrt{\frac{3}{2}} e^{-\phi_4}}{4 \pi }, \quad  {x=a,b,c}.
\end{align}
To calculate gaugino masses $M_{1,2,3}$ respectively for ${\rm U}(1)_Y$, ${\rm SU}(2)_L$ and ${\rm SU}(3)_c$ gauge groups, we first compute $M_{a,b,c}$ using \eqref{gaugino-masses} as,
\begin{align}
M_a &= \frac{\sqrt{3} }{2} m_{3/2} \left(e^{-i \gamma_2} \Theta_3+e^{-i \gamma_3} \Theta_2\right),\nonumber \\
M_b &= \frac{\sqrt{3} }{2} m_{3/2} \left(e^{-i \gamma_2} \Theta_2-e^{-i \gamma_4} \Theta_4\right),\nonumber \\
M_c &= \frac{\sqrt{3} }{2} m_{3/2} \left(e^{-i \gamma_1} \Theta_1+e^{-i \gamma_3} \Theta_3\right).
\end{align}
Bino mass parameter $M_Y$ \eqref{Bino-mass} is then computed as,
\begin{align}
M_Y &= \frac{1}{f_Y}(\frac{2}{3}f_aM_a+f_cM_c) \nonumber\\
&=\frac{3 \sqrt{3}}{10}m_{3/2}\left(e^{-i \gamma _1} \Theta _1+\frac{1}{5} e^{-i \gamma _2} \Theta _2+\frac{1}{2} e^{-i \gamma _3} \Theta _3\right).
\end{align}
Therefore, the gaugino masses for ${\rm U}(1)_Y$, ${\rm SU}(2)_L$ and ${\rm SU}(3)_c$ gauge groups are,
\begin{align}\label{Gauginos123}
M_1 &\equiv M_Y = \frac{3 \sqrt{3}}{10}m_{3/2}\left(e^{-i \gamma _1} \Theta _1+\frac{1}{5} e^{-i \gamma _2} \Theta _2+\frac{1}{2} e^{-i \gamma _3} \Theta _3\right), \nonumber\\
M_2 &\equiv M_b = \frac{\sqrt{3}}{2} m_{3/2} \left(e^{-i \gamma_2} \Theta_2-e^{-i \gamma_4} \Theta_4\right),\nonumber\\
M_3 &\equiv M_a = \frac{\sqrt{3}}{2} m_{3/2} \left(e^{-i \gamma_2} \Theta_3+e^{-i \gamma_3} \Theta_2\right).
\end{align}

\begin{table}[t] \footnotesize
\renewcommand{\arraystretch}{1.0}
\caption{The angles (in multiples of $\pi$) with respect to the orientifold plane
made by the cycle wrapped by each stack of D-branes
on each of the three two-tori.}
\label{Angles3}
\begin{center}
\begin{tabular}{|c|c|c|c|}\hline
 & $\theta_1$
&  $\theta_2$  & $\theta_3$ \\
\hline
$a$ & $-1/2$  & $\ 1/4$  & $ 1/4$\\
$b$ & $\ 1/4$  & $\ 0 $  & $-1/4$\\
$c$ & $-1/4$ & $\ 1/2$   & $-1/4$\\
\hline
\end{tabular}
\end{center}
\end{table}

We now require the angles, the differences of angles and their first and second order derivatives with respect to the moduli to compute the trilinear coupling and the sleptons mass-squared. In table~\ref{Angles3} we show the angles made by the cycle wrapped by each stack D6 branes with respect to the orientifold plane in multiples of
$\pi$,
\begin{align}\label{eqn:Angle}
\theta_{x}^i &= \frac{1}{\pi}\tan^{-1}\left(\frac{2^{-\beta_i}l_x^i\chi_i}{n_x^i}\right).
\end{align}

The differences of the angles are,
\begin{align}\label{angle-diff}
\theta_{y}^{i} -\theta_{x}^{i} \equiv \theta_{xy}^{i} =\left(
\begin{array}{ccc}
 \{0,0,0\} & \left\{\frac{3}{4},-\frac{1}{4},-\frac{1}{2}\right\} & \left\{\frac{1}{4},\frac{1}{4},-\frac{1}{2}\right\} \\
 \left\{-\frac{3}{4},\frac{1}{4},\frac{1}{2}\right\} & \{0,0,0\} & \left\{-\frac{1}{2},\frac{1}{2},0\right\} \\
 \left\{-\frac{1}{4},-\frac{1}{4},\frac{1}{2}\right\} & \left\{\frac{1}{2},-\frac{1}{2},0\right\} & \{0,0,0\} \\
\end{array}
\right).
\end{align}

To account for the negative angle differences we make use of the $\sigma_{xy}^{i}$ function \eqref{sigmaK} which is $-1$ only for negative angle difference and $+1$
otherwise,
\begin{align}\label{sigma3}
\sigma_{xy}^{i} \equiv (-1)^{1-H(\theta_{xy}^{i})} & = \left(
\begin{array}{ccc}
 \{1,1,1\} & \{-1,-1,1\} & \{1,-1,1\} \\
 \{1,1,-1\} & \{1,1,1\} & \{1,-1,1\} \\
 \{-1,1,-1\} & \{-1,1,1\} & \{1,1,1\} \\
\end{array}
\right),
\end{align}
where $H(x)$ is the unit step function. And the function $\eta_{xy}$ can thus be defined by taking the product on the torus index $i$ as,
\begin{equation}\label{eta3}
\eta_{xy}\equiv  \prod_i \sigma_{xy}^i = \left(
\begin{array}{ccc}
 1 & 1 & -1 \\
 -1 & 1 & -1 \\
 1 & -1 & 1 \\
\end{array}
\right).
\end{equation}
Using above defined $\sigma_{xy}^{i}$ and $\eta_{xy}^{i}$, we can readily write the four cases of functions $\Psi(\theta_{xy})$ defined in \eqref{eqn:Psi1} and
\eqref{eqn:Psi2} into a single expression as,
\begin{align}\label{Psi}
\Psi(\theta^j_{xy}) &= \eta_{xy}\left( \frac{1}{2} \psi^{(0)}(\sigma_{xy}^{i}\theta^j_{xy})+\frac{1}{2}\psi^{(0)}(1-\sigma_{xy}^{i}\theta^j_{xy})+\gamma_E-\log(t^j+\bar
t^j)\right) ,
\end{align}
where $\psi^{(0)}(z)$ is called the digamma function defined as the derivative of the logarithm of the gamma function. The successive derivatives of the $\log\Gamma (z)$
yield the polygamma function $\psi^{(n)}(z)$ as,
\begin{align}\label{polygamma}
\psi^{(n-1)}(z) = \frac{d^{(n)}}{dz^{(n)}}\log\Gamma(z) ,
\end{align}
with the following properties,
\begin{align}\label{property}
\frac{d }{dz}\psi^{(0)}(\pm z) &= \pm \psi^{(1)}(\pm z), \nonumber\\
\frac{d }{dz}\psi^{(0)}(1 \pm z) &= \pm \psi^{(1)}(1 \pm z).
\end{align}
Similarly, the derivative $\Psi'(\theta^j_{xy}) = \frac{d\Psi(\theta^j_{xy})}{d \theta^j_{xy}}$ can be expressed succinctly as,
\begin{align}\label{DPsi}
\Psi'(\theta^j_{xy}) & = \eta_{xy}\sigma_{xy}^{i} \left( \frac{1}{2} \psi^{(1)}(\sigma_{xy}^{i}\theta^j_{xy})+\frac{1}{2}\psi^{(1)}(1-\sigma_{xy}^{i}\theta^j_{xy})\right),
\end{align}
where we have utilized the property \eqref{property} and have neglected the contribution of the $t$-moduli.

Lastly, by making use of appropriate Kronecker deltas and defining $u^4\equiv s$, we can express the various cases of the first and second derivatives of the angles as,
\begin{align}\label{derivative-angles}
{\theta}^{i,m}_{xy} \equiv (u^m+\bar u^m)\,\frac{\partial \theta^i_{xy}}{\partial u^m} = (-1)^{\delta _{m,4}} (-1)^{\delta _{i,j}}\frac{\sin (2 \pi  \theta^i)}{4\pi }
\bigg|^x_y , \nonumber \\
i={1,2,3}; \quad m={1,2,3,4.}
\end{align}
\begin{align}
{\theta}^{i,mn}_{xy} &\equiv (u^m+\bar u^m)(u^n+\bar u^n)\,\frac{\partial^2 \theta^i_{xy}}{\partial u^m\partial\bar u^n} \nonumber \\
&= \delta _{m,n}\frac{\sin (4 \pi  \theta^i) +(-1)^{(1-\delta _{4,m}) (1-\delta _{i,m})}4 \sin (2 \pi  \theta^i)}{16\pi } \bigg|^x_y \nonumber \\
&\quad + (1-\delta _{m,n}) (-1)^{(1-\delta _{4,m}) (1-\delta _{4,n}) (\delta _{i,m}+\delta _{i,n})} (-1)^{1-\delta _{i,m}-\delta _{i,n}} \frac{\sin (4 \pi  \theta^i)}{16\pi
} \bigg|^x_y ,\nonumber \\
& \qquad \qquad \qquad \qquad\qquad\qquad\qquad\qquad i={1,2,3}; \quad m,n={1,2,3,4}.
\end{align}

Substituting above results in \eqref{tri-coupling} and ignoring the CP-violating phases $\gamma^i$, we obtain the following the trilinear couplings,
\begin{align}\label{A0}
A_0 & \equiv A_{abc}= A_{acb} = \frac{\sqrt{3} m_{3/2}}{4 \pi } \Bigg[-\Theta_1 \left(2 \pi +2 \gamma_{E} +\psi ^{(0)}\left(\frac{1}{4}\right)+\psi ^{(0)}\left(\frac{3}{4}\right)\right)\nonumber\\
&\quad+\Theta_2 \left(-2 \pi +2 \gamma_{E} +\psi ^{(0)}\left(\frac{1}{4}\right)+\psi ^{(0)}\left(\frac{3}{4}\right)\right) \nonumber\\
&\quad+\Theta_3 \left(2 \gamma_{E} +\psi ^{(0)}\left(\frac{1}{4}\right)+\psi ^{(0)}\left(\frac{3}{4}\right)\right)-\Theta_4 \left(2 \gamma_{E} +\psi ^{(0)}\left(\frac{1}{4}\right)+\psi
^{(0)}\left(\frac{3}{4}\right)\right)\Bigg].
\end{align}
Ignoring the CP-violating phases $\gamma^i$, the gaugino masses, trilinear coupling and sleptons and squarks mass-squared \eqref{slepton-mass} parameters are obtained as,
\begin{align}
M_1 & =m_{3/2} (0.519615 \Theta _1+0.34641 \Theta _2+0.866025 \Theta _3) ,\nonumber\\
M_2 & =m_{3/2} (0.866025 \Theta _2-0.866025 \Theta _4) ,\nonumber\\
M_3 & =m_{3/2} ( 0.866025 \Theta _2+0.866025 \Theta _3) ,\nonumber\\
A_0 &= m_{3/2}(-0.292797 \Theta_1-1.43925 \Theta_2-0.573228 \Theta_3 +0.573228 \Theta_4) ,\nonumber\\
m^2_{ab} & \equiv m^2_{L} = m_{3/2}{}^2 \Big(1-2.02977 \Theta_1{}^2+0.75 \Theta_1 \Theta_2-1.5 \Theta_1 \Theta_4-0.0440466 \Theta_2{}^2-1.5 \Theta_2
\Theta_3 \nonumber\\
&\quad\quad\quad +0.286907 \Theta_3{}^2+0.75 \Theta_3 \Theta_4+0.286907 \Theta_4{}^2\Big) ,\nonumber\\
m^2_{ac}& \equiv m^2_{R} = m_{3/2}{}^2 \Big(1-0.0880932 \Theta_1{}^2-1.5 \Theta_1 \Theta_2+0.75 \Theta_1 \Theta_3+0.75 \Theta_1 \Theta_4-0.0880932
\Theta_2{}^2\nonumber\\
&\quad\quad\quad+0.75 \Theta_2 \Theta_3+0.75 \Theta_2 \Theta_4-0.419047 \Theta_3{}^2-1.5 \Theta_3 \Theta_4-2.40477 \Theta_4{}^2\Big) .
\end{align}
All above results are subject to the constraint,
\begin{align}\label{constraint}
\sum_{i=1}^4 \Theta_i^2=1.
\end{align}

\bibliographystyle{JHEP}

\begin{thebibliography}{10}

\bibitem{Chen:2007zu}
C.-M. Chen, T.~Li, V.~E. Mayes and D.~V. Nanopoulos, \emph{{Towards realistic
  supersymmetric spectra and Yukawa textures from intersecting branes}},
  \href{https://doi.org/10.1103/PhysRevD.77.125023}{\emph{Phys. Rev. D}
  {\bfseries 77} (2008) 125023}
  [\href{https://arxiv.org/abs/0711.0396}{{\ttfamily 0711.0396}}].

\bibitem{Witten:2002ei}
E.~Witten, \emph{{Quest for unification}},  in \emph{{10th International
  Conference on Supersymmetry and Unification of Fundamental Interactions
  (SUSY02)}}, pp.~604--610, 7, 2002,
  \href{https://arxiv.org/abs/hep-ph/0207124}{{\ttfamily hep-ph/0207124}}.

\bibitem{Chamoun:2003pf}
N.~Chamoun, S.~Khalil and E.~Lashin, \emph{{Fermion masses and mixing in
  intersecting branes scenarios}},
  \href{https://doi.org/10.1103/PhysRevD.69.095011}{\emph{Phys. Rev. D}
  {\bfseries 69} (2004) 095011}
  [\href{https://arxiv.org/abs/hep-ph/0309169}{{\ttfamily hep-ph/0309169}}].

\bibitem{Higaki:2005ie}
T.~Higaki, N.~Kitazawa, T.~Kobayashi and K.-j. Takahashi, \emph{{Flavor
  structure and coupling selection rule from intersecting D-branes}},
  \href{https://doi.org/10.1103/PhysRevD.72.086003}{\emph{Phys. Rev. D}
  {\bfseries 72} (2005) 086003}
  [\href{https://arxiv.org/abs/hep-th/0504019}{{\ttfamily hep-th/0504019}}].

\bibitem{Cvetic:2004ui}
M.~Cvetic, T.~Li and T.~Liu, \emph{{Supersymmetric Pati-Salam models from
  intersecting D6-branes: A Road to the standard model}},
  \href{https://doi.org/10.1016/j.nuclphysb.2004.07.036}{\emph{Nucl. Phys. B}
  {\bfseries 698} (2004) 163}
  [\href{https://arxiv.org/abs/hep-th/0403061}{{\ttfamily hep-th/0403061}}].

\bibitem{Li:2019nvi}
T.~Li, A.~Mansha and R.~Sun, \emph{{Revisiting the supersymmetric
  Pati\textendash{}Salam models from intersecting D6-branes}},
  \href{https://doi.org/10.1140/epjc/s10052-021-08839-w}{\emph{Eur. Phys. J. C}
  {\bfseries 81} (2021) 82} [\href{https://arxiv.org/abs/1910.04530}{{\ttfamily
  1910.04530}}].

\bibitem{Gimon:1996rq}
E.~G. Gimon and J.~Polchinski, \emph{{Consistency conditions for orientifolds
  and D-manifolds}},
  \href{https://doi.org/10.1103/PhysRevD.54.1667}{\emph{Phys. Rev. D}
  {\bfseries 54} (1996) 1667}
  [\href{https://arxiv.org/abs/hep-th/9601038}{{\ttfamily hep-th/9601038}}].

\bibitem{Green:1984sg}
M.~B. Green and J.~H. Schwarz, \emph{{Anomaly Cancellation in Supersymmetric
  D=10 Gauge Theory and Superstring Theory}},
  \href{https://doi.org/10.1016/0370-2693(84)91565-X}{\emph{Phys. Lett. B}
  {\bfseries 149} (1984) 117}.

\bibitem{Witten:1998cd}
E.~Witten, \emph{{D-branes and K-theory}},
  \href{https://doi.org/10.1088/1126-6708/1998/12/019}{\emph{JHEP} {\bfseries
  12} (1998) 019} [\href{https://arxiv.org/abs/hep-th/9810188}{{\ttfamily
  hep-th/9810188}}].

\bibitem{Cascales:2003zp}
J.~F.~G. Cascales and A.~M. Uranga, \emph{{Chiral 4d string vacua with D branes
  and NSNS and RR fluxes}},
  \href{https://doi.org/10.1088/1126-6708/2003/05/011}{\emph{JHEP} {\bfseries
  05} (2003) 011} [\href{https://arxiv.org/abs/hep-th/0303024}{{\ttfamily
  hep-th/0303024}}].

\bibitem{Marchesano:2004yq}
F.~Marchesano and G.~Shiu, \emph{{MSSM vacua from flux compactifications}},
  \href{https://doi.org/10.1103/PhysRevD.71.011701}{\emph{Phys. Rev. D}
  {\bfseries 71} (2005) 011701}
  [\href{https://arxiv.org/abs/hep-th/0408059}{{\ttfamily hep-th/0408059}}].

\bibitem{Marchesano:2004xz}
F.~Marchesano and G.~Shiu, \emph{{Building MSSM flux vacua}},
  \href{https://doi.org/10.1088/1126-6708/2004/11/041}{\emph{JHEP} {\bfseries
  11} (2004) 041} [\href{https://arxiv.org/abs/hep-th/0409132}{{\ttfamily
  hep-th/0409132}}].

\bibitem{Uranga:2000xp}
A.~M. Uranga, \emph{{D-brane probes, RR tadpole cancellation and K-theory
  charge}}, \href{https://doi.org/10.1016/S0550-3213(00)00787-2}{\emph{Nucl.
  Phys. B} {\bfseries 598} (2001) 225}
  [\href{https://arxiv.org/abs/hep-th/0011048}{{\ttfamily hep-th/0011048}}].

\bibitem{Chen:2006gd}
C.-M. Chen, T.~Li and D.~V. Nanopoulos, \emph{{Type IIA Pati-Salam flux
  vacua}}, \href{https://doi.org/10.1016/j.nuclphysb.2006.01.039}{\emph{Nucl.
  Phys. B} {\bfseries 740} (2006) 79}
  [\href{https://arxiv.org/abs/hep-th/0601064}{{\ttfamily hep-th/0601064}}].

\bibitem{Cvetic:2004nk}
M.~Cvetic, P.~Langacker, T.-j. Li and T.~Liu, \emph{{D6-brane splitting on type
  IIA orientifolds}},
  \href{https://doi.org/10.1016/j.nuclphysb.2004.12.028}{\emph{Nucl. Phys. B}
  {\bfseries 709} (2005) 241}
  [\href{https://arxiv.org/abs/hep-th/0407178}{{\ttfamily hep-th/0407178}}].

\bibitem{Cvetic:2007ku}
M.~Cvetic, R.~Richter and T.~Weigand, \emph{{Computation of D-brane instanton
  induced superpotential couplings: Majorana masses from string theory}},
  \href{https://doi.org/10.1103/PhysRevD.76.086002}{\emph{Phys. Rev. D}
  {\bfseries 76} (2007) 086002}
  [\href{https://arxiv.org/abs/hep-th/0703028}{{\ttfamily hep-th/0703028}}].

\bibitem{Blumenhagen:2006xt}
R.~Blumenhagen, M.~Cvetic and T.~Weigand, \emph{{Spacetime instanton
  corrections in 4D string vacua: The Seesaw mechanism for D-Brane models}},
  \href{https://doi.org/10.1016/j.nuclphysb.2007.02.016}{\emph{Nucl. Phys. B}
  {\bfseries 771} (2007) 113}
  [\href{https://arxiv.org/abs/hep-th/0609191}{{\ttfamily hep-th/0609191}}].

\bibitem{Haack:2006cy}
M.~Haack, D.~Krefl, D.~Lust, A.~Van~Proeyen and M.~Zagermann, \emph{{Gaugino
  Condensates and D-terms from D7-branes}},
  \href{https://doi.org/10.1088/1126-6708/2007/01/078}{\emph{JHEP} {\bfseries
  01} (2007) 078} [\href{https://arxiv.org/abs/hep-th/0609211}{{\ttfamily
  hep-th/0609211}}].

\bibitem{Florea:2006si}
B.~Florea, S.~Kachru, J.~McGreevy and N.~Saulina, \emph{{Stringy Instantons and
  Quiver Gauge Theories}},
  \href{https://doi.org/10.1088/1126-6708/2007/05/024}{\emph{JHEP} {\bfseries
  05} (2007) 024} [\href{https://arxiv.org/abs/hep-th/0610003}{{\ttfamily
  hep-th/0610003}}].

\bibitem{Cremmer:1982en}
E.~Cremmer, S.~Ferrara, L.~Girardello and A.~Van~Proeyen, \emph{{Yang-Mills
  Theories with Local Supersymmetry: Lagrangian, Transformation Laws and
  SuperHiggs Effect}},
  \href{https://doi.org/10.1016/0550-3213(83)90679-X}{\emph{Nucl. Phys. B}
  {\bfseries 212} (1983) 413}.

\bibitem{Lust:2004cx}
D.~Lust, P.~Mayr, R.~Richter and S.~Stieberger, \emph{{Scattering of gauge,
  matter, and moduli fields from intersecting branes}},
  \href{https://doi.org/10.1016/j.nuclphysb.2004.06.052}{\emph{Nucl. Phys. B}
  {\bfseries 696} (2004) 205}
  [\href{https://arxiv.org/abs/hep-th/0404134}{{\ttfamily hep-th/0404134}}].

\bibitem{Kane:2004hm}
G.~L. Kane, P.~Kumar, J.~D. Lykken and T.~T. Wang, \emph{{Some phenomenology of
  intersecting D-brane models}},
  \href{https://doi.org/10.1103/PhysRevD.71.115017}{\emph{Phys. Rev. D}
  {\bfseries 71} (2005) 115017}
  [\href{https://arxiv.org/abs/hep-ph/0411125}{{\ttfamily hep-ph/0411125}}].

\bibitem{Blumenhagen:2006ci}
R.~Blumenhagen, B.~Kors, D.~Lust and S.~Stieberger, \emph{{Four-dimensional
  String Compactifications with D-Branes, Orientifolds and Fluxes}},
  \href{https://doi.org/10.1016/j.physrep.2007.04.003}{\emph{Phys. Rept.}
  {\bfseries 445} (2007) 1}
  [\href{https://arxiv.org/abs/hep-th/0610327}{{\ttfamily hep-th/0610327}}].

\bibitem{Klebanov:2003my}
I.~R. Klebanov and E.~Witten, \emph{{Proton decay in intersecting D-brane
  models}}, \href{https://doi.org/10.1016/S0550-3213(03)00410-3}{\emph{Nucl.
  Phys. B} {\bfseries 664} (2003) 3}
  [\href{https://arxiv.org/abs/hep-th/0304079}{{\ttfamily hep-th/0304079}}].

\bibitem{Blumenhagen:2003jy}
R.~Blumenhagen, D.~Lust and S.~Stieberger, \emph{{Gauge unification in
  supersymmetric intersecting brane worlds}},
  \href{https://doi.org/10.1088/1126-6708/2003/07/036}{\emph{JHEP} {\bfseries
  07} (2003) 036} [\href{https://arxiv.org/abs/hep-th/0305146}{{\ttfamily
  hep-th/0305146}}].

\bibitem{Ibanez:2001nd}
L.~E. Ibanez, F.~Marchesano and R.~Rabadan, \emph{{Getting just the standard
  model at intersecting branes}},
  \href{https://doi.org/10.1088/1126-6708/2001/11/002}{\emph{JHEP} {\bfseries
  11} (2001) 002} [\href{https://arxiv.org/abs/hep-th/0105155}{{\ttfamily
  hep-th/0105155}}].

\bibitem{Blumenhagen:2005tn}
R.~Blumenhagen, M.~Cvetic, F.~Marchesano and G.~Shiu, \emph{{Chiral D-brane
  models with frozen open string moduli}},
  \href{https://doi.org/10.1088/1126-6708/2005/03/050}{\emph{JHEP} {\bfseries
  03} (2005) 050} [\href{https://arxiv.org/abs/hep-th/0502095}{{\ttfamily
  hep-th/0502095}}].

\bibitem{Font:2004cx}
A.~Font and L.~E. Ibanez, \emph{{SUSY-breaking soft terms in a MSSM magnetized
  D7-brane model}},
  \href{https://doi.org/10.1088/1126-6708/2005/03/040}{\emph{JHEP} {\bfseries
  03} (2005) 040} [\href{https://arxiv.org/abs/hep-th/0412150}{{\ttfamily
  hep-th/0412150}}].

\bibitem{Cvetic:2003ch}
M.~Cvetic and I.~Papadimitriou, \emph{{Conformal field theory couplings for
  intersecting D-branes on orientifolds}},
  \href{https://doi.org/10.1103/PhysRevD.70.029903}{\emph{Phys. Rev. D}
  {\bfseries 68} (2003) 046001}
  [\href{https://arxiv.org/abs/hep-th/0303083}{{\ttfamily hep-th/0303083}}].

\bibitem{Kawamura:1996ex}
Y.~Kawamura, T.~Kobayashi and T.~Komatsu, \emph{{Specific scalar mass relations
  in SU(3) x SU(2) x U(1) orbifold model}},
  \href{https://doi.org/10.1016/S0370-2693(97)00364-X}{\emph{Phys. Lett. B}
  {\bfseries 400} (1997) 284}
  [\href{https://arxiv.org/abs/hep-ph/9609462}{{\ttfamily hep-ph/9609462}}].

\bibitem{Komargodski:2009pc}
Z.~Komargodski and N.~Seiberg, \emph{{Comments on the Fayet-Iliopoulos Term in
  Field Theory and Supergravity}},
  \href{https://doi.org/10.1088/1126-6708/2009/06/007}{\emph{JHEP} {\bfseries
  06} (2009) 007} [\href{https://arxiv.org/abs/0904.1159}{{\ttfamily
  0904.1159}}].

\bibitem{Brignole:1997dp}
A.~Brignole, L.~E. Ibanez and C.~Munoz, \emph{{Soft supersymmetry breaking
  terms from supergravity and superstring models}},
  \href{https://doi.org/10.1142/9789812839657_0003}{\emph{Adv. Ser. Direct.
  High Energy Phys.} {\bfseries 18} (1998) 125}
  [\href{https://arxiv.org/abs/hep-ph/9707209}{{\ttfamily hep-ph/9707209}}].

\bibitem{Brignole:1993dj}
A.~Brignole, L.~E. Ibanez and C.~Munoz, \emph{{Towards a theory of soft terms
  for the supersymmetric Standard Model}},
  \href{https://doi.org/10.1016/0550-3213(94)00068-9}{\emph{Nucl. Phys. B}
  {\bfseries 422} (1994) 125}
  [\href{https://arxiv.org/abs/hep-ph/9308271}{{\ttfamily hep-ph/9308271}}].

\bibitem{Aldazabal:2000cn}
G.~Aldazabal, S.~Franco, L.~E. Ibanez, R.~Rabadan and A.~M. Uranga,
  \emph{{Intersecting brane worlds}},
  \href{https://doi.org/10.1088/1126-6708/2001/02/047}{\emph{JHEP} {\bfseries
  02} (2001) 047} [\href{https://arxiv.org/abs/hep-ph/0011132}{{\ttfamily
  hep-ph/0011132}}].

\bibitem{Cremades:2003qj}
D.~Cremades, L.~E. Ibanez and F.~Marchesano, \emph{{Yukawa couplings in
  intersecting D-brane models}},
  \href{https://doi.org/10.1088/1126-6708/2003/07/038}{\emph{JHEP} {\bfseries
  07} (2003) 038} [\href{https://arxiv.org/abs/hep-th/0302105}{{\ttfamily
  hep-th/0302105}}].

\bibitem{Fusaoka:1998vc}
H.~Fusaoka and Y.~Koide, \emph{{Updated estimate of running quark masses}},
  \href{https://doi.org/10.1103/PhysRevD.57.3986}{\emph{Phys. Rev. D}
  {\bfseries 57} (1998) 3986}
  [\href{https://arxiv.org/abs/hep-ph/9712201}{{\ttfamily hep-ph/9712201}}].

\bibitem{Ross:2007az}
G.~Ross and M.~Serna, \emph{{Unification and fermion mass structure}},
  \href{https://doi.org/10.1016/j.physletb.2008.05.014}{\emph{Phys. Lett. B}
  {\bfseries 664} (2008) 97} [\href{https://arxiv.org/abs/0704.1248}{{\ttfamily
  0704.1248}}].

\bibitem{Abel:2003yx}
S.~A. Abel and A.~W. Owen, \emph{{N point amplitudes in intersecting brane
  models}}, \href{https://doi.org/10.1016/j.nuclphysb.2003.11.032}{\emph{Nucl.
  Phys. B} {\bfseries 682} (2004) 183}
  [\href{https://arxiv.org/abs/hep-th/0310257}{{\ttfamily hep-th/0310257}}].

\bibitem{Abel:2003vv}
S.~A. Abel and A.~W. Owen, \emph{{Interactions in intersecting brane models}},
  \href{https://doi.org/10.1016/S0550-3213(03)00370-5}{\emph{Nucl. Phys. B}
  {\bfseries 663} (2003) 197}
  [\href{https://arxiv.org/abs/hep-th/0303124}{{\ttfamily hep-th/0303124}}].

\bibitem{Chen:2008rx}
C.-M. Chen, T.~Li, V.~E. Mayes and D.~V. Nanopoulos, \emph{{Yukawa Corrections
  from Four-Point Functions in Intersecting D6-Brane Models}},
  \href{https://doi.org/10.1103/PhysRevD.78.105015}{\emph{Phys. Rev. D}
  {\bfseries 78} (2008) 105015}
  [\href{https://arxiv.org/abs/0807.4216}{{\ttfamily 0807.4216}}].

\bibitem{Mayes:2013bda}
V.~E. Mayes, \emph{{Universal Soft Terms in the MSSM on D-branes}},
  \href{https://doi.org/10.1016/j.nuclphysb.2013.10.016}{\emph{Nucl. Phys. B}
  {\bfseries 877} (2013) 401}
  [\href{https://arxiv.org/abs/1305.2842}{{\ttfamily 1305.2842}}].

\bibitem{Mayes:2019isy}
V.~E. Mayes, \emph{{All Fermion Masses and Mixings in an Intersecting D-brane
  World}}, \href{https://doi.org/10.1016/j.nuclphysb.2019.114848}{\emph{Nucl.
  Phys. B} {\bfseries 950} (2020) 114848}
  [\href{https://arxiv.org/abs/1902.00983}{{\ttfamily 1902.00983}}].

\bibitem{Gemmill:2019kxr}
J.~Gemmill, E.~Howington and V.~E. Mayes, \emph{{Fitting neutrino masses in a
  realistic intersecting D-braneworld}},
  \href{https://doi.org/10.1103/PhysRevD.100.115048}{\emph{Phys. Rev. D}
  {\bfseries 100} (2019) 115048}
  [\href{https://arxiv.org/abs/1907.07106}{{\ttfamily 1907.07106}}].

\bibitem{Li:2014xqa}
T.~Li, D.~V. Nanopoulos, S.~Raza and X.-C. Wang, \emph{{A Realistic
  Intersecting D6-Brane Model after the First LHC Run}},
  \href{https://doi.org/10.1007/JHEP08(2014)128}{\emph{JHEP} {\bfseries 08}
  (2014) 128} [\href{https://arxiv.org/abs/1406.5574}{{\ttfamily 1406.5574}}].

\end{thebibliography}

\providecommand{\href}[2]{#2}\begingroup\raggedright\endgroup

\end{document}